\documentclass[aps,onecolumn,superscriptaddress,floatfix,nofootinbib]{revtex4}

\usepackage[pdftex]{graphicx}
\usepackage{amsmath,amssymb}
\usepackage{bm}
\usepackage{color}
\usepackage[normalem]{ulem}
\usepackage[usenames,dvipsnames]{xcolor}
\usepackage{eufrak}
\usepackage{bbm}

\newcommand{\MP}[1]{{\color{black}{{#1}}}}
\newcommand{\WEU}[1]{{\color{black}{{#1}}}}

\begin{document}

\title{Shape-dependent guidance of active Janus particles by chemically patterned surfaces}

\author{W. E. Uspal}
\email[Corresponding author: ]{uspal@is.mpg.de}
\author{M. N. Popescu}
\affiliation{Max-Planck-Institut f\"{u}r Intelligente Systeme, Heisenbergstr. 3, 
D-70569 Stuttgart, Germany}
\affiliation{IV. Institut f\"ur Theoretische Physik, Universit{\"a}t Stuttgart, 
Pfaffenwaldring 57, D-70569 Stuttgart, Germany}

\author{M. Tasinkevych}
\affiliation{Centro de F\'\i sica Te\'orica e Computacional, 
Departamento de F\'\i sica, Faculdade 
de Ci\^encias, Universidade de Lisboa, Campo Grande P-1749-016 Lisboa, Portugal
}

\author{S. Dietrich}
\affiliation{Max-Planck-Institut f\"{u}r Intelligente Systeme, Heisenbergstr. 3, 
D-70569 Stuttgart, Germany}
\affiliation{IV. Institut f\"ur Theoretische Physik, Universit{\"a}t Stuttgart, 
Pfaffenwaldring 57, D-70569 Stuttgart, Germany}

\date{\today}

\begin{abstract}
Self-phoretic chemically active Janus particles move by inducing -- via non-equilibrium chemical 
reactions occurring on their surfaces -- changes in the chemical composition of the solution
in which they are immersed. This process leads to gradients in chemical composition along the 
surface of the particle, as well as along any nearby boundaries, including solid walls. Chemical 
gradients along a wall can give rise to chemi-osmosis, i.e., the gradients 
drive surface flows which, in turn, drive flow in the volume of the solution. 
This bulk flow couples back to the particle, and thus contributes to its self-motility. Since 
chemi-osmosis strongly depends on the molecular interactions between the diffusing molecular 
species and the wall, the response flow induced and experienced by a particle encodes information
about any chemical patterning of the wall. Here, we extend previous studies on 
self-phoresis of a sphere 
near a chemically patterned wall to the case of particles with rod-like, elongated shape. We focus 
our analysis on the new phenomenology potentially emerging from the coupling  -- which is 
inoperative for a spherical shape -- of the elongated particle to the
strain rate tensor of the chemi-osmotic flow. Via detailed numerical calculations, we show that
the dynamics of a rod-like particle exhibits a novel ``edge-following'' steady state: the particle
translates along the edge of a chemical step at a steady distance from the step and with a steady
orientation. Moreover, within a certain range of system parameters, the edge-following state
co-exists with a ``docking'' state (the particle stops at the step, oriented perpendicular to the 
step edge), i.e.,
a bistable dynamics occurs. These findings are rationalized as a consequence of the competition
between the fluid vorticity and the rate of strain by using analytical
theory based on the point-particle approximation which captures quasi-quantitatively the dynamics 
of the system.
\end{abstract}

\maketitle

\section{Introduction}

Over the past two decades, significant effort has been invested in the development of 
nano- and micro-particles that can propel themselves through an aqueous 
environment \MP{\cite{ebbens10,gompper15,bechinger16}}. These synthetic swimmers have myriad potential 
applications in, e.g., 
cell sorting and manipulation \cite{baraban12b,solovev12}, micromanufacturing \cite{goodrich17}, 
and the assembly of dynamic and programmable materials \cite{stenhammar16}. For these and other 
applications, a longstanding challenge is to endow the synthetic swimmers with a semblance of 
``intelligence'', i.e., the ability to autonomously sense the local environment (e.g., ambient flow 
fields \cite{palacci15,uspal15b}, or their position with respect to confining 
surfaces \cite{das15,simmchen16}) and to respond according to their design, \MP{i.e., to exhibit some 
form of ``taxis.'' For example, recent studies have reported, experimentally and theoretically, on 
the gravitaxis of asymmetrically shaped colloids \cite{hagen14}, as well as on phototaxis of light-activated 
colloids in a spatially modulated light intensity landscape \cite{lozano16,geiseler17}.}

Among artificial swimmers, catalytically active particles constitute a large and important 
class \cite{paxton06,moran17,aubret17}. These particles typically move by 
self-phoresis, i.e.,  by consuming molecular ``fuel'' -- available in the surrounding 
solution -- they generate gradients of one or more thermodynamic variables (e.g., chemical 
composition or electrical potential) in the solution. These gradients, in conjunction with the 
interaction of the particle surface with the molecules of the solution, drive flow of the 
solution past the particle, powering directed motion \cite{golestanian05,golestanian07}.  For 
example, bimetallic Au/Pt rods consume hydrogen peroxide and release charged molecules 
into the solution, and hence move by self-electrophoresis \cite{paxton04}. Janus spheres with an 
inert core and a platinum cap, when exposed to hydrogen peroxide, move by self-diffusiophoresis  
\cite{howse07} or, as proposed recently, by a more complex mechanism involving self-
electrophoretic contributions \cite{brown14,ebbens14,brown17}. In addition to spheres and rods, 
catalytically active particles with more complex shapes have been fabricated, e.g., dimers 
of two connected particles \cite{valadares10,palacci15}, ``matchsticks,'' \cite{morgan14} and 
microstars \cite{simmchen17}. 

Self-phoretic particles exhibit complex dynamics when moving in the vicinity of a confining boundary. 
As with other swimmers, e.g., those moving due to mechanical surface deformations, self-phoretic 
particles create flow fields that extend into the surrounding solution, are reflected from nearby 
boundaries, and couple back to the particles \cite{spagnolie12}. These hydrodynamic interactions 
can lead to bound states of motion in which the swimming particle always remains in the vicinity 
of the boundary \cite{ishimoto13}. Notably, an additional level of complexity arises for self-phoretic 
particles, because the thermodynamic gradients that drive particle motion will  also be modified 
by confining boundaries. For instance, if the particle creates gradients of chemical composition, 
a solid boundary will present a no-flux boundary condition for the number density fields of the 
various molecular species diffusing in the solution. In a previous study, we have shown that a 
no-flux boundary condition, in combination with hydrodynamic interactions, can lead to novel 
bound states, such as motionless ``hovering'' of a catalytic Janus sphere above a planar solid 
boundary (or  ``wall'') \cite{uspal15}.

\WEU{All of these phenomena are enriched even further if the boundary itself is ``responsive,'' i.e., 
if thermodynamic gradients along the boundary translate into tangential surfaces stresses driving 
flow of the solution.} Indeed, in analogy with phoresis, ``chemi-osmotic'' surface flows can occur if 
gradients in chemical composition occur along the surface of a solid boundary 
\cite{anderson89,derjaguin:47}. For a catalytic particle, chemi-osmotic surface flows, \textit{locally} 
induced by the particle,  will drive flow in the bulk solution, coupling back to the particle. Recently, 
the propulsion of catalytic ``surfers'' near a solid boundary was attributed to such an effect 
\cite{palacci13}. Significantly, chemi-osmotic flows depend on the material identity of the solid 
boundary. This naturally leads to the question of whether, by patterning the chemical composition of 
the responsive wall, one can guide the motion of a catalytically active particle. Recently, 
we investigated the interplay of chemi-osmosis and self-diffusiophoresis for a catalytically active 
Janus sphere above a patterned wall \cite{uspal16,popescu17}. By analytical and numerical calculations 
we demonstrated that a sphere that swims \textit{away} from its catalytic cap can ``dock'' at a 
chemical step between two materials. In contrast, a sphere that swims \textit{towards} its cap can 
align with and follow a chemical stripe. We identified the physical mechanisms driving these 
behaviors. One key finding is that a chemically patterned wall can create fluid \textit{vorticity} in 
the bulk solution, and that this wall-induced vorticity completely accounts for rotation of the 
spherical particle.

\begin{figure}[t]
\includegraphics[scale=0.8]{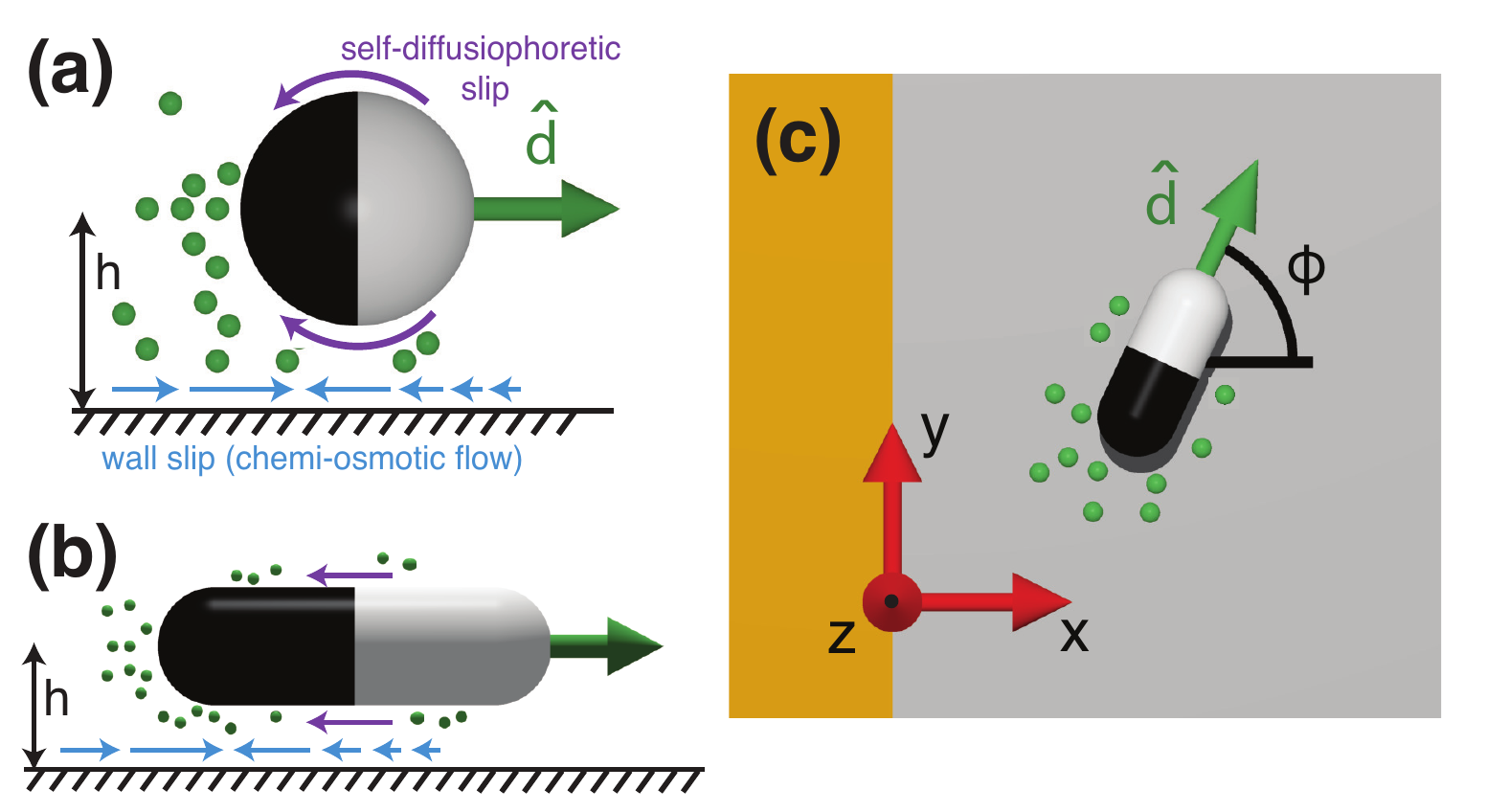}
\caption{\label{fig:schematic} (a, b) Schematic views of a Janus sphere and a Janus spherocylinder, respectively, at a distance $h$ from a solid planar wall. Each particle produces solute molecules (green spheres) over the area of its catalytic cap (black). The effective interaction between the solute and the particle surface drives a self-diffusiophoretic flow (purple arrows) at the surface of the particle. Similarly, the solute drives chemi-osmotic 
surface flow at the wall (blue arrows). The orientation vector $\mathbf{\hat{d}}$ is defined to point from the catalytic pole of the particle to the inert pole. For repulsive 
interactions between the solute molecules and the wall (which corresponds to the direction of the flow shown here), the surface flow converges towards the region of maximum density of solute. This surface flow entrains flow in the bulk solution (not shown). Near the point of convergence, due to fluid incompressibility the bulk flow must lift off the surface into the volume of the solution, coupling back to the particle. (c) Top-down schematic view of a spherocylinder near a chemically patterned surface. In this case, the pattern is a chemical step, as defined by Eq. \ref{eq:step_surface_mob}. The orientation vector $\mathbf{\hat{d}}$ and the $\hat{x}$ axis form the angle $\phi$.}
\end{figure}

The orientational dynamics of a solid particle in a spatially varying ambient flow field is a classical problem in microhydrodynamics \cite{kim1991}. A natural starting point is to perform a Taylor expansion of the flow velocity with respect to the particle centroid (i.e., its geometrical center). At first order in such an expansion, the vorticity can be mathematically identified with the antisymmetric component of the first derivative of the flow velocity. The symmetric component of this tensor, the so-called \textit{rate of strain} tensor, does not contribute to the rotation of a sphere. 
However, it does contribute to rotation of a non-spherical or rod-like particle, such as a spheroid or spherocylinder. Consequently, the orientational dynamics of a rod-like particle in a spatially varying flow field is typically much more complex than for a sphere. For instance, a sphere in a linear shear flow will simply rotate around the axis of vorticity at a rate proportional to the vorticity, while any other solid body of revolution will execute a complex three-dimensional periodic motion known as a Jeffery orbit \cite{jeffery22,bretherton62}.

These previous findings naturally raise the question of how a rod-like catalytically active particle will behave near a chemically patterned wall. The present study addresses 
this question. Via detailed numerical calculations, we show that a rod-like particle, as compared with a sphere, exhibits significantly richer dynamics near a patterned wall. We find a new ``edge-following'' steady state, in which a rod-like particle moves along the edge of a chemical step at a steady distance from the step and with a steady orientation. Moreover, within a certain range of system parameters, the edge-following state co-exists with a docking state, i.e., the dynamics is bistable. We develop approximate analytical expressions that isolate the contributions of fluid vorticity and the rate of strain to particle rotation. We show that the rich phenomenology of rod-like particles is indeed a 
consequence of the competition between fluid vorticity and rate of strain.

\section{Formulation of the model}

We consider a catalytic Janus particle in the vicinity of a hard planar wall. The particle 
is a solid body of revolution with axially symmetric surface chemistry. Two 
examples of particles of this kind -- a sphere and a spherocylinder -- are shown 
schematically in Fig. \ref{fig:schematic}.  We assume a stationary reference frame in 
which the wall is located 
at $z = 0$ and the instantaneous position of the particle centroid is 
$\mathbf{x}_{p}(t) = (x_p(t), y_p(t), z_p(t))$; for brevity of notations, in the following we shall 
indicate the dependence on time only if it is explicitly required. We denote by $\mathbf{x}$ a point 
in the liquid surrounding the particle. Over half of its surface (i.e., the catalytic ``cap''), the particle 
consumes ``fuel'' molecules, taken to be present in abundance in the solution, and 
emits product molecules (``solute''). We additionally assume that the reaction accounting 
for the conversion of the fuel molecules into solute molecules is such that the system 
operates within the so-called regime of reaction limited kinetics (see, e.g., Ref. \cite{oshanin17} and references therein). Under these 
assumptions, the chemical activity of the particle can be modeled simply as a release of 
solute molecules at a constant rate $\kappa$ (areal number density per time) over half 
of its surface (the catalytic ``cap''), which corresponds to a so-called 
``constant-flux'' model \cite{golestanian05}. 
The orientation of the particle is defined by a unit vector $\mathbf{\hat{d}}$, which 
is aligned with the axis of symmetry of the particle, and points from the catalytic cap to the inert ``face'' of the particle. The solute diffuses in the surrounding liquid with a diffusion constant $D$. We define the P\'{e}clet number $Pe \equiv U_{0} R_{1}/D$, where $U_0$ and $R_1$ are the characteristic velocity and the characteristic length scale of the particle, respectively. If the P\'{e}clet number is small, the solute number density field 
$c(\mathbf{x})$ is quasi-static, and can be taken to obey the Laplace equation $\nabla^{2} c = 0$. The wall and the inert face are taken to be impenetrable to solute, so that the number density field is subject to the boundary conditions  $- D [\mathbf{\hat{n}} \cdot \nabla c] = \kappa$ on the catalytic cap, and $- D [\mathbf{\hat{n}} \cdot \nabla c] = 0$ on the inert face and on the wall, where the normal $\mathbf{\hat{n}}$ is defined to point into the liquid. Additionally, the number density field is required to attain a fixed value $c^{\infty}$ 
far away from the particle, i.e., $c(|\mathbf{x} - \mathbf{x}_{p}| \rightarrow \infty) = c^{\infty}$.

The suspending solution, consisting of solvent, fuel, and solute molecules, is 
an incompressible Newtonian fluid with dynamic viscosity $\eta$ and mass density 
$\rho$.  We use the classical theory of charge neutral diffusiophoresis in order to model 
how concentration gradients drive particle motion through the liquid \cite{anderson89}. 
Within this framework, the interaction between the solute molecules and a confining 
surface (i.e., the wall or one of the particle faces) drives a surface flow which is 
described by an effective slip velocity $\mathbf{v}_{s}(\mathbf{x}_{s}) = -b(\mathbf{x}_{s}) 
\nabla_{||} c(\mathbf{x})|_{\mathbf{x}=\mathbf{x}_{s}}$, where $\nabla_{||} \equiv 
(\mathcal{I} - \mathbf{\hat{n}} \mathbf{\hat{n}}) \cdot \nabla$, $\mathcal{I}$ is the 
identity tensor, and $\mathbf{x}_{s}$ denotes a position on that surface. The so-called 
surface mobility $b\mathbf(\mathbf{x}_{s})$ is a material-dependent parameter which
encodes the molecular interaction between the solute and that surface. We consider 
small Reynolds numbers $Re \equiv \rho U_0 R_{1} / \eta$, so that the hydrodynamic 
flow $\mathbf{u}(\mathbf{x})$ and the pressure $P(\mathbf{x})$ of the solution obey 
the Stokes equation $-\nabla P + \eta \nabla^{2} \mathbf{u} = 0$ and the incompressibility 
condition $\nabla \cdot \mathbf{u} = 0$.  The fluid velocity obeys the boundary conditions 
$\mathbf{u}\lvert_{wall}=\mathbf{v}_{s}(\mathbf{x}_{s})$ on the wall, and 
$\mathbf{u}\lvert_{part}=\mathbf{U}+\bm{\Omega}\times(\mathbf{x}_{s}-\mathbf{x}_{p})+
\mathbf{v}_{s}(\mathbf{x}_{s})$ on the particle surface; note that the dependence on 
$\mathbf{x}_{s}$ accounts also for the, in general, distinct surface mobilities. 
Additionally, the fluid is quiescent far away from the particle, so that $\mathbf{u}(|
\mathbf{x} - \mathbf{x}_{p}| \rightarrow \infty) = 0$.  Here, $\mathbf{U}$ and $\bm{\Omega}$ 
are unknown translational and angular velocities, respectively, of the particle.  In order 
to close the system of equations for $\mathbf{U}$ and $\bm{\Omega}$, we invoke the force balance, $\int \, \bm{\sigma} \cdot \mathbf{\hat{n}} \, dS = -\mathbf{F}^{ext}$, and the torque balance, $\int \, (\mathbf{x} - \mathbf{x}_{p}) \times \bm{\sigma} \cdot \mathbf{\hat{n}} \, dS = -\bm{\tau}^{ext}$, for the particle. 
The fluid stress tensor is $\bm{\sigma} = -P \mathcal{I} + \eta \left[ \nabla \mathbf{u} + \nabla^{T} \mathbf{u} \right]$. Here, $\mathbf{F}^{ext}$ and $\bm{\tau}^{ext}$ are the net external force and net external torque, respectively; these can represent (for instance) particle buoyancy, bottom-heaviness, and any external forces and torques of constraint. The above two integrals are taken over the surface of the particle.

The above considerations completely specify how to determine the instantaneous 
particle velocities $\mathbf{U}$ and $\bm{\Omega}$ as functions of the instantaneous 
position $\mathbf{x}_{p}$ and the orientation $\mathbf{\hat{d}}$ of the particle. This 
scheme includes various contributions to the particle velocities, such as from slip on 
the surface of the particle (self-diffusiophoresis); slip on the surface of the wall 
(chemi-osmosis), which drives hydrodynamic flow in the volume of the solution; and 
external forces and torques. At this point, we make some conceptual distinctions which 
will prove to be helpful later. We note that the Stokes equations are linear. Therefore, the 
full problem can be split into three subproblems which isolate the \MP{individual contributions 
of wall slip (\textbf{ws}), self-diffusiophoresis (\textbf{sd}), and external forces and 
torques (\textbf{ext}).}  In subproblem 
\textbf{(sd)}, we take the surface velocity on the wall to be zero, i.e., $\mathbf{v}_{s}
(\mathbf{x}_{s})\lvert_{wall} \,= 0$, as well as $\mathbf{F}^{ext} = 0$ and $\bm{\tau}^{ext} = 
0$, but otherwise the subproblem is posed as given above.  This subproblem can be solved for 
$\mathbf{U}^{sd}$ and $\bm{\Omega}^{sd}$. Similarly, in subproblem \textbf{(ws)}, we take 
$\mathbf{v}_{s}(\mathbf{x}_{s})\lvert_{part} \,= 0$, $\mathbf{F}^{ext} = 0$ and $\bm{\tau}^{ext} = 0$. 
We obtain $\mathbf{U}^{ws}$ and $\bm{\Omega}^{ws}$ by solving subproblem \textbf{(ws)}. Finally, 
if there are external forces and torques, we consider a subproblem \textbf{(ext)}, in which 
$\mathbf{v}_{s}(\mathbf{x}_{s})\lvert_{wall} \,= 0$ and 
$\mathbf{v}_{s}(\mathbf{x}_{s})\lvert_{part} \,= 0$, and which can be solved for $\mathbf{U}^{ext}$ 
and $\bm{\Omega}^{ext}$.  We superpose the individual solutions to  obtain the total 
particle velocities $\mathbf{U} = \mathbf{U}^{sd} + \mathbf{U}^{ws} + \mathbf{U}^{ext}$ and 
$\bm{\Omega} = \bm{\Omega}^{sd} +  \bm{\Omega}^{ws} + \bm{\Omega}^{ext}$. 

For a sphere, the self-diffusiophoretic contributions $\mathbf{U}^{sd}$ and $\bm{\Omega}^{sd}$ had 
been considered in Ref. \cite{uspal15}. The very same framework developed in that study can be 
applied to spherocylinders. Additionally, the contributions $\mathbf{U}^{ext}$ and $\bm{\Omega}^{ext}$  
of the external forces and torques can be solved with standard methods (e.g., see 
Ref. \cite{simmchen16}.) In the remainder of the 
present study, we focus on obtaining the contributions from chemi-osmotic wall slip.  

At this stage we make some simplifying assumptions. We take the particle height $z_p(t)$ to 
be a fixed quantity $z_p(t) = h$, and assume that $\mathbf{\hat{d}}$ is always oriented within 
the plane of the wall, i.e., $\mathbf{\hat{d}} \cdot \hat{z} = 0$, where $\hat{z}$ is 
the unit vector in $z$-direction. In this case the orientation is completely 
described by the angle $\phi$, as shown in Fig. \ref{fig:schematic}(c). For a 
spherocylindrical particle, these assumptions can be viewed as realistic. Catalytic 
particles are typically heavy, and they will sediment to the bottom surface of a 
bounding container. Concerning the restriction on $\mathbf{\hat{d}}$, if $h$ is small 
(compared to the pole-to-pole length of the particle), ``rocking'' of the particle orientation 
$\mathbf{\hat{d}}$ out of the $xy$ plane is restricted to small angles by the presence of the 
hard wall; this is apparent from inspecting Fig. \ref{fig:schematic}(b). For spheres, the restriction 
on $\mathbf{\hat{d}}$ could be imposed by an external magnetic torque. \MP{Furthermore, we note that 
for heavy colloids the surface-to-surface distance is expected to be small (when compared to 
the radius of the particle): it is set mainly by the balance between the weight of the particle and 
the DLVO repulsion. If the vertical component of the osmotic flow is not very large, which is 
expected to be the case for our system, it is a reasonable assumption that the change in height due 
to the particle being dragged by the vertical component of the osmotic flow at the location of the 
particle is small compared to $h$, i.e., $h/R$ remains practically unchanged.} (For 
a further discussion of this assumption in relation to spheres, the reader is invited to peruse 
Ref. \cite{uspal16}.)  We note that based on these assumptions in Ref. 
\cite{uspal15} it was shown via symmetry considerations that $\mathbf{U}^{sd}$ and $\bm{\Omega}^{sd}$ 
are functions of $h$ and $\mathbf{\hat{d}} \cdot \hat{z}$ only, and that ${\Omega}^{sd}_{z} = 0$. 
Since we take $h$ to be constant and 
$\mathbf{\hat{d}} \cdot \hat{z} = 0$, one finds 
$\mathbf{U}^{sd} = U^{sd} \, \mathbf{\hat{d}}$ and ${\Omega}^{sd}_{z} = 0$. In the present 
study, we treat $U^{sd}$ as an input parameter; however, for a given particle geometry and 
surface chemistry, it can be calculated via the methods outlined in Ref. \cite{uspal15}.

The position and orientation of the particle evolve according to the equations
\begin{equation}
\label{eq:particle_dynamics}
\begin{split}
\dot{x}_{p} &= U^{sd} \cos(\phi) + U_{x}^{ws}(x_p, y_p, \phi), \\
\dot{y}_{p} &= U^{sd} \sin(\phi) + U_{y}^{ws}(x_p, y_p, \phi),~\mathrm{and}\\
\dot{\phi} &= \Omega^{ws}_{z}(x_p, y_p, \phi),
\end{split}
\end{equation}
where the over-dot denotes the time derivative of the corresponding observable. These equations can be integrated to obtain complete particle trajectories \footnote{Note that 
we have assumed that external forces and torques do not contribute to the in-plane motion of the particle. In some cases, this assumption represents an additional approximation. For instance, for a sphere subject to a magnetic torque of constraint, there is a contribution to the in-plane velocity of the particle which is due to wall-induced translational-rotational coupling. As discussed at length in Ref. \cite{popescu17}, this contribution is negligible for the particle-wall separations considered here.}. 

We consider two types of surface patterns: a ``chemical step'' and a ``chemical stripe.'' For the chemical step, the surface mobility $b(\mathbf{x}_{s})$ is
\begin{equation}
\label{eq:step_surface_mob}
b(\mathbf{x}_{s}) = 
\begin{cases}
b_{w}^{l},~ x_s < 0,\\
b_{w}^{r},~ x_s \geq 0,\hspace*{1cm}
\end{cases} 
\end{equation}
where $b_{w}^{l}$ and $b_{w}^{r}$ hold on the \textit{l}eft and \textit{r}ight hand sides 
of the \textit{w}all, respectively. The chemical stripe is defined by
\begin{equation}
\label{eq:stripe_surface_mob}
b(\mathbf{x}_{s}) = 
\begin{cases}
b_{w},~ x_s < -W,\\
b_{w}^{c},~ -W \leq x_s \leq W, \\ 
b_{w},~ x_s > W, \hspace*{1cm}
\end{cases} 
\end{equation}
i.e., the \textit{c}entral stripe has $b=b_{w}^{c}$, while $b=b_{w}$ on the rest of the wall. Both the chemical step and the chemical stripe have translational symmetry in the $\hat{y}$-direction. Therefore, the dependence of $U_{x}^{ws}$, $U_{y}^{ws}$, and $\Omega_{z}^{ws}$ on ${y}_p$ drops out of Eq. (\ref{eq:particle_dynamics}), and the dynamics is completely described by two degrees of freedom, ${x}_{p}$ and $\phi$. 

We express lengths in units of $R_{1}$, where $2 R_{1}$ is the largest dimension of the particle. The solute density field has a characteristic number density $\kappa R_{1}/D$. Combining this with the surface mobility gives a characteristic velocity $U_0 = |b_0 \kappa/D|$, where $b_0$ is a characteristic value of the surface mobility. For the chemical step, we choose $b_0 = b_w^{l}$, and for the stripe, we choose $b_0 = b_w^{c}$. \WEU{We express $U^{sd}$ in units of $U_0$ so that, for a given particle geometry and height, the dimensionless parameter $U^{sd}/U_{0}$ characterizes the relative strengths of the self-diffusiophoretic and the chemi-osmotic contributions to particle motility.} For reasons of space, in this study we restrict our considerations to 
$U^{sd} > 0$, i.e., to a particle which tends to move away from its catalytic cap via self-diffusiophoresis. This ``inert-forward'' direction of motion is more typically seen in experiments than the ``catalyst-forward'' direction of motion. (Nevertheless, a catalyst-forward particle can also exhibit an interesting behavior near a chemically patterned surface, as we demonstrated for spheres in Ref. \cite{uspal16}.) An additional dimensionless parameter characterizes the surface: $b_{w}^{r}/b_{w}^{l}$ for the chemical step and $b_{w}/b_{w}^{c}$ 
for the stripe, respectively.\newline

\section{Numerical results and discussion}

In order to numerically solve the problem posed in the previous section, we resort to the 
so-called Boundary Element Method (BEM) \cite{uspal15}. The effect of a non-spherical shape 
is revealed by considering two particle geometries: a Janus sphere 
and a spherocylinder, each of them half covered by catalyst. These geometries are shown 
schematically in Fig. \ref{fig:schematic}(a) and (b). The sphere has a radius $R$, and 
the height $h$ of the sphere is fixed as $h/R = 1.1$. The major axis of the spherocylinder 
has a tip-to-tip length $2c$, and the radius of the cylindrical part of the particle is 
chosen to be  $a = 0.4 c$.  The height of the spherocylinder is fixed as 
$h = 0.6 c$. 

\begin{figure}[!htb]
\includegraphics[scale=0.45]{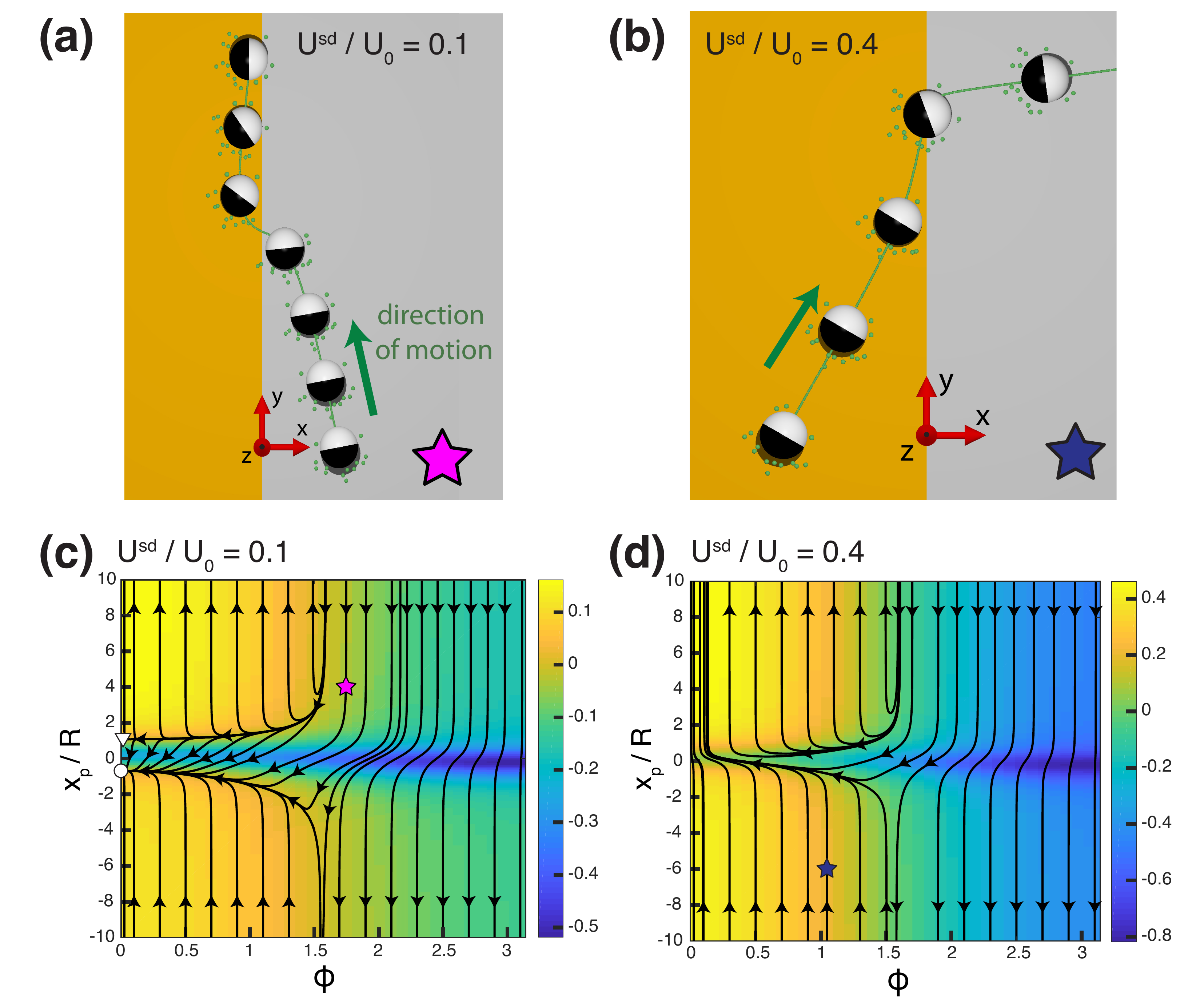}
\caption{\label{fig:sphere_traj_portraits} (a-b) Trajectories of a 
self-diffusiophoretic sphere, half-covered by catalyst, at height $h/R = 1.1$. 
For (a) and (b), the self-propulsion velocity is $U^{sd}/U_{0} = 0.1$ and 
$U^{sd}/U_{0} = 0.4$, respectively. Star symbols match the trajectories to the 
initial conditions shown in the phase portraits in (c) and (d), respectively. 
(c-d) Phase portraits for a half-covered sphere at a height $h/R = 1.1$ for two 
values of the self-diffusiophoretic velocity $U^{sd}$. The white circle indicates 
an attractor, whereas the white triangle indicates a saddle point. The color 
field encodes the component $U_{x}$ of the particle velocity in $\hat{x}$ direction.}
\end{figure}
The motion of a sphere near a chemical step had previously been studied in 
Ref. \cite{uspal16}. We briefly revisit those findings here, choosing $b_{w}^{r}/b_{w}^{l} = 4$ 
and $b_{w}^{l} < 0$. Therefore, both regions of the substrate have a repulsive 
effective interaction with the solute molecules, but the region to the right of the step ($x_p > 0$) has a stronger repulsive interaction. At low self-diffusiophoretic velocity $U^{sd}$, and depending on the initial orientation $\phi$ and distance $x_{p}$ from 
the step, the sphere can spontaneously ``dock'' at the step. In the docking state, the 
particle is trapped and motionless near the step (i.e., $|x_p/R| \ll 1$), with the 
orientation vector $\mathbf{\hat{d}}$ pointing towards the region where the substrate 
is more repulsive to the solute. A particle trajectory with $U^{sd}/U_0 = 0.1$, which 
is attracted to the docking state, is illustrated in Fig. \ref{fig:sphere_traj_portraits}(a). The initial condition of this trajectory is matched by the magenta star symbol to Fig. \ref{fig:sphere_traj_portraits}(c), which is a  phase portrait showing the time evolution of any initial state $(x_{p}, \phi)$. From the phase portrait, it is clear that if the particle approaches the edge from the left ($x_p < 0$), the particle will dock for nearly any angle of incidence $0 \leq \phi < \pi/2$. 
If the particle approaches from the right ($x_p > 0$), 
the angle of incidence must be oblique ($\phi \gtrsim \pi/2$) for the particle to 
dock; otherwise, the particle will pass over the step. At higher speeds $U^{sd}$, 
the particle always passes over the step. Figure \ref{fig:sphere_traj_portraits}(b) 
shows the trajectory of a particle with $U^{sd}/U_0 = 0.4$; the corresponding phase 
portrait is shown in Fig. \ref{fig:sphere_traj_portraits}(d).  Interestingly, if the 
particle passes from the left $(x_p < 0$, $\phi < \pi/2$), there is a narrow range of 
exit angles $\phi \ll 1$ for a broad range of angles of approach $0 \leq \phi < \pi/2$. 
In other words, in this case the chemical step operates 
as an ``angular focusing'' device. If the particle passes from the right ($x_p > 0$, 
$\phi > \pi/2$), the deviation of the exit angle from the initial angle is modest. 

\begin{figure}[!htb]
\includegraphics[scale=0.5]{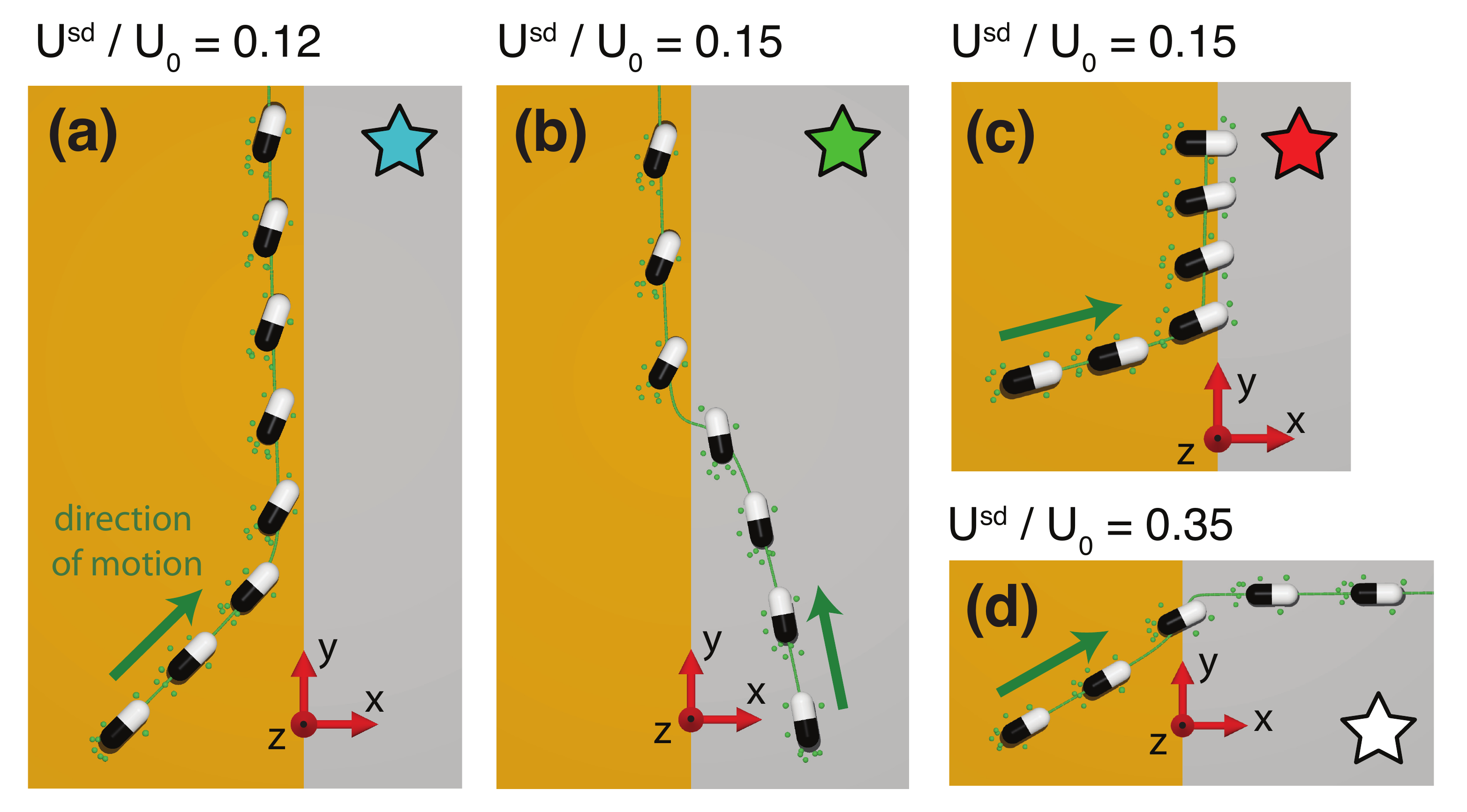}
\caption{\label{fig:traj_spherocyl} Trajectories of a half-covered spherocylinder with height 
$h/c = 0.6$ near a chemical step. For (a), the self-propulsion velocity is 
$U^{sd}/U_{0} = 0.12$. For (b)-(c), the speed is $U^{sd}/U_{0} = 0.15$, and for 
(d) it is $U^{sd}/U_{0} = 0.35$. The star symbols match the trajectories to the 
initial conditions in Fig. \ref{fig:phase_portrait_spherocyl}.}
\end{figure}
\begin{figure}[!htb]
\includegraphics[scale=0.5]{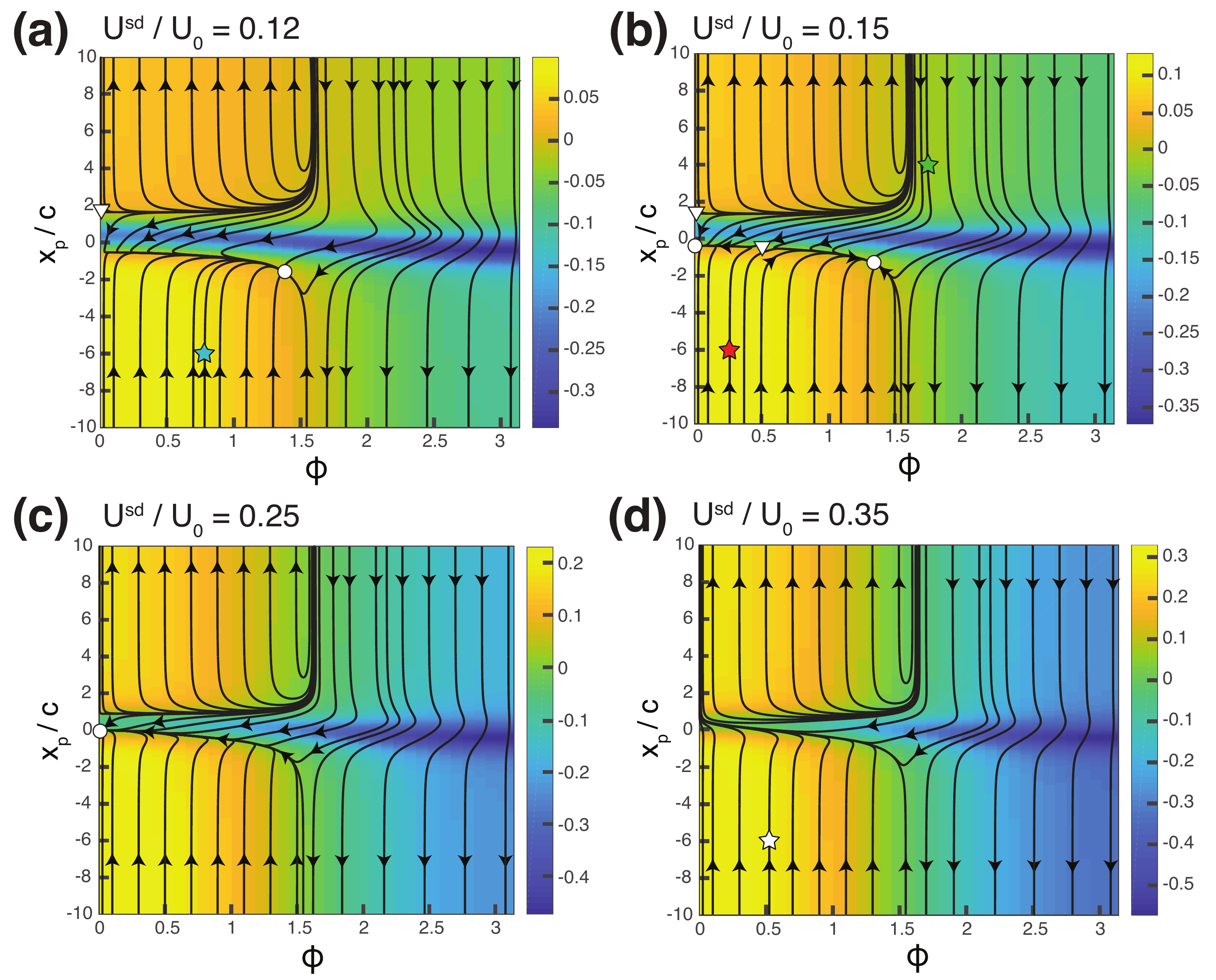}
\caption{\label{fig:phase_portrait_spherocyl}Phase portraits of a half-covered spherocylinder 
near a chemical step for various self-diffusiophoretic velocities $U^{sd}$. The particle is at 
a height $h/c = 0.6$. The star symbols indicate the initial conditions corresponding to the 
trajectories shown in Fig. \ref{fig:traj_spherocyl}. White circles 
indicate attractors, whereas white triangles indicate saddle points. The color field 
encodes the component $U_{x}$ of the particle velocity in $\hat{x}$ direction.}
\end{figure}

Now we consider the spherocylindrical particle near the same chemical step 
(Fig. \ref{fig:traj_spherocyl}). We find that the phenomenology is considerably enriched by 
the elongated shape. At low speed $U^{sd}$, there is a stable ``edge-following'' state in 
which the particle remains at a fixed distance $|x_p/c| \ll 1$ and angle $\phi \lesssim 
\pi/2$. Accordingly, the particle orientation $\mathbf{\hat{d}}$ aligns almost parallel 
to the step edge, but tilted slightly towards the more repulsive region of the substrate, 
and the particle moves steadily along the edge. A phase portrait for $U^{sd}/U_0 = 0.12$ 
is shown in Fig. \ref{fig:phase_portrait_spherocyl}(a); the white circle represents the 
edge-following attractor. The cyan star in the phase portrait gives the initial condition 
for the trajectory in Fig. \ref{fig:traj_spherocyl}(a). At a slightly higher speed 
$U^{sd}/U_0 = 0.15$, the edge-following state (Fig. \ref{fig:traj_spherocyl}(b)) co-exists 
with a stable docking state (Fig. \ref{fig:traj_spherocyl}(c)), i.e., the dynamics is 
bistable. For the speed $U^{sd}/U_0 = 0.15$ edge-following and docking trajectories 
are illustrated in Figs. \ref{fig:traj_spherocyl}(b) and (c); the complete phase 
portrait corresponding to $U^{sd}/U_0 = 0.15$ is shown in Fig. 
\ref{fig:phase_portrait_spherocyl}(b). The bistability occurs within a finite range of speed 
values $U^{sd}$. As $U^{sd}$ is increased within this range, the attractors and the saddle 
points of the phase diagram move in the phase space $(\phi, x_{p}/c)$. Within this 
process, at the upper threshold of this range, the saddle point and the edge-following state 
collide and annihilate, leaving only the docking state. The docking state persists for a 
certain range of high speeds $U^{sd}$. In Fig. \ref{fig:phase_portrait_spherocyl}(c), we show 
the phase portrait for $U^{sd}/U_0 = 0.25$. As $U^{sd}$ is increased even further, the docking 
state attractor also collides with a saddle point and annihilates. Therefore, for very high 
speed $U^{sd}$, the spherocylinder passes over the step for any inital angle with 
$\mathbf{\hat{d}}$ pointing towards the step. A corresponding phase portrait for 
$U^{sd}/U_{0} = 0.35$ is shown in Fig. \ref{fig:phase_portrait_spherocyl}(d). As in the 
case of the sphere, a spherocylinder passing from the left has a narrow band of exit angles, 
while a spherocylinder passing from the right shows only modest deviations of the exit angles 
$\phi$ from the incoming ones. Such a trajectory, which passes from the left, is shown in 
Fig. \ref{fig:traj_spherocyl}(d).

We note that we have obtained a very similar phenomenology for a prolate spheroid, i.e., 
edge-following at low speed $U^{sd}$, bistable edge-following and docking at 
intermediate speeds $U^{sd}$, etc. This agreement indicates that the enrichment of the 
phenomenology as compared to that for a sphere is a generic consequence of non-spherical 
shapes, rather than being specific for spherocylinders. We omit 
these calculations in order to avoid an overly repetitive presentation. Additionally, we 
note that for both spheres and spherocylinders the phenomenology we have discussed above 
is observed for a range of other values of the surface mobility contrast 
$(b_{w}^{r}/b_{w}^{l} - 1)$ which are of $\mathcal{O}(1)$. 

\begin{figure}
\includegraphics[scale=0.9]{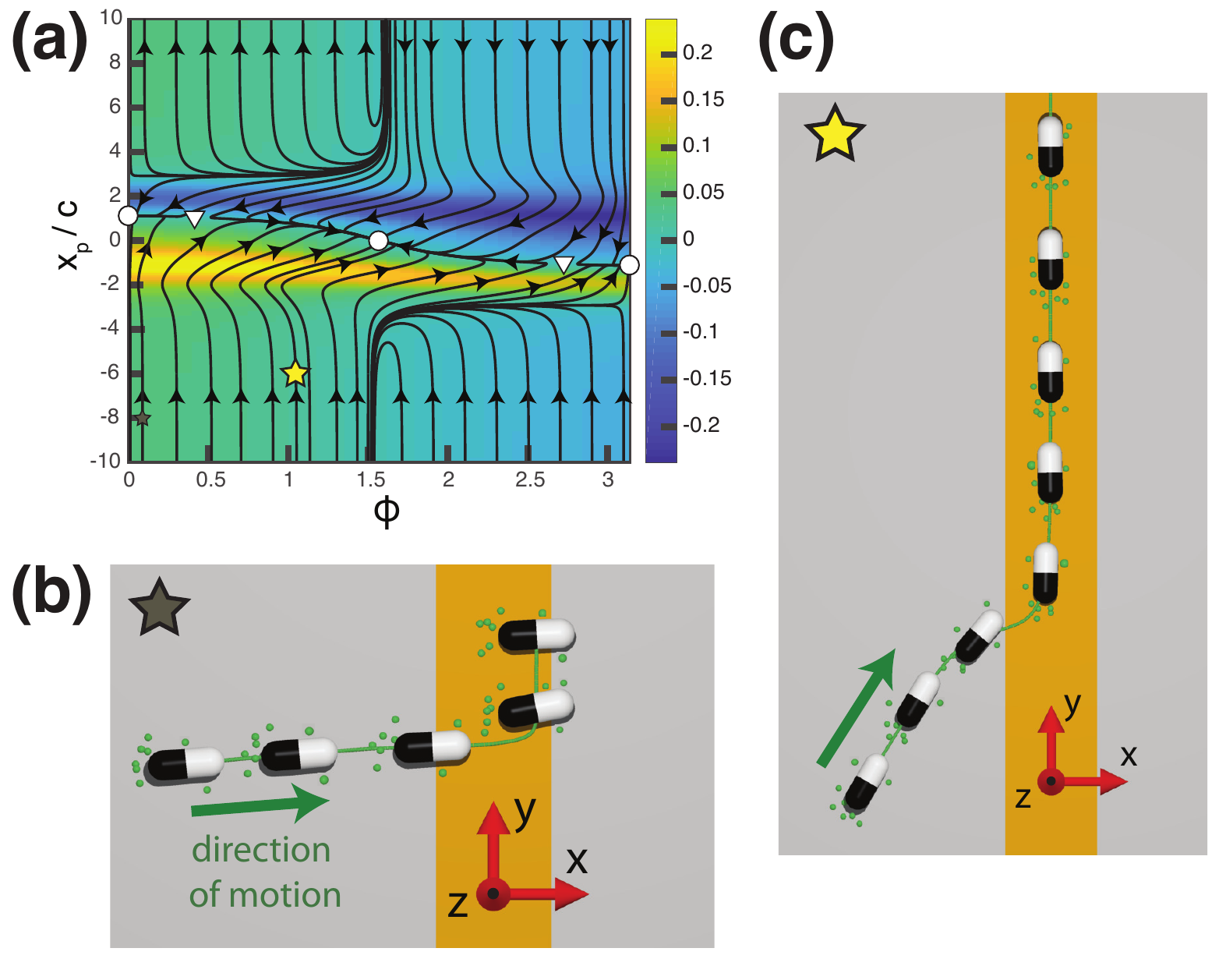}
\caption{\label{fig:stripe_figure} (a) Phase portrait for a half-covered spherocylinder 
near a chemical stripe. The particle is at height $h/c = 0.6$ and has a 
self-diffusiophoretic velocity $U^{sd}/U_0 = 0.1$. Star symbols indicate initial 
conditions corresponding to the trajectories shown in panels (b) and (c). White circles 
indicate attractors, whereas white triangles indicate saddle points. The color field 
in (a) encodes the component $U_x$ of the particle velocity in $\hat{x}$ direction. 
Panels (b) and (c) show a docking and a stripe-following trajectory, respectively, of 
a spherocylinder near a chemical stripe. }
\end{figure}
We briefly discuss the behavior of a rod-like particle near a chemical stripe. As one 
might expect, if the stripe width $2W$ is large, the dynamics near the stripe resembles 
the one near two isolated chemical steps. However, as the width is decreased and 
becomes comparable to the particle size, the effects of the two edges start to interfere with 
each other. We consider a stripe with $2W = 3 c$, $b_{w}/b_{w}^{c} = 3$, and $b_{w}^{c} < 0$. 
Both the sides and the center of the stripe have a repulsive effective interaction with 
the solute, but this interaction is stronger for the sides than for the center.  In 
Fig. \ref{fig:stripe_figure}(a), we show the phase portrait for a spherocylinder near 
this stripe with $U^{sd}/U_{0} = 0.1$. The dynamics is multistable. There are two docking 
states, associated with the two edges. But instead of two edge-following states, we obtain 
a ``stripe-following'' attractor in which the particle is positioned at the center of the 
stripe ($x_p = 0$) and is perfectly aligned with it ($\phi = \pi/2$). Docking and stripe-
following trajectories are shown in Fig. \ref{fig:stripe_figure}(b) and (c).  For 
comparison, in previous studies we had obtained a stripe-following state for Janus spheres, 
but only for $U^{sd} < 0$, i.e., for a particle which tends to move towards its 
catalytic face. 

\begin{figure}
\includegraphics[scale=0.45]{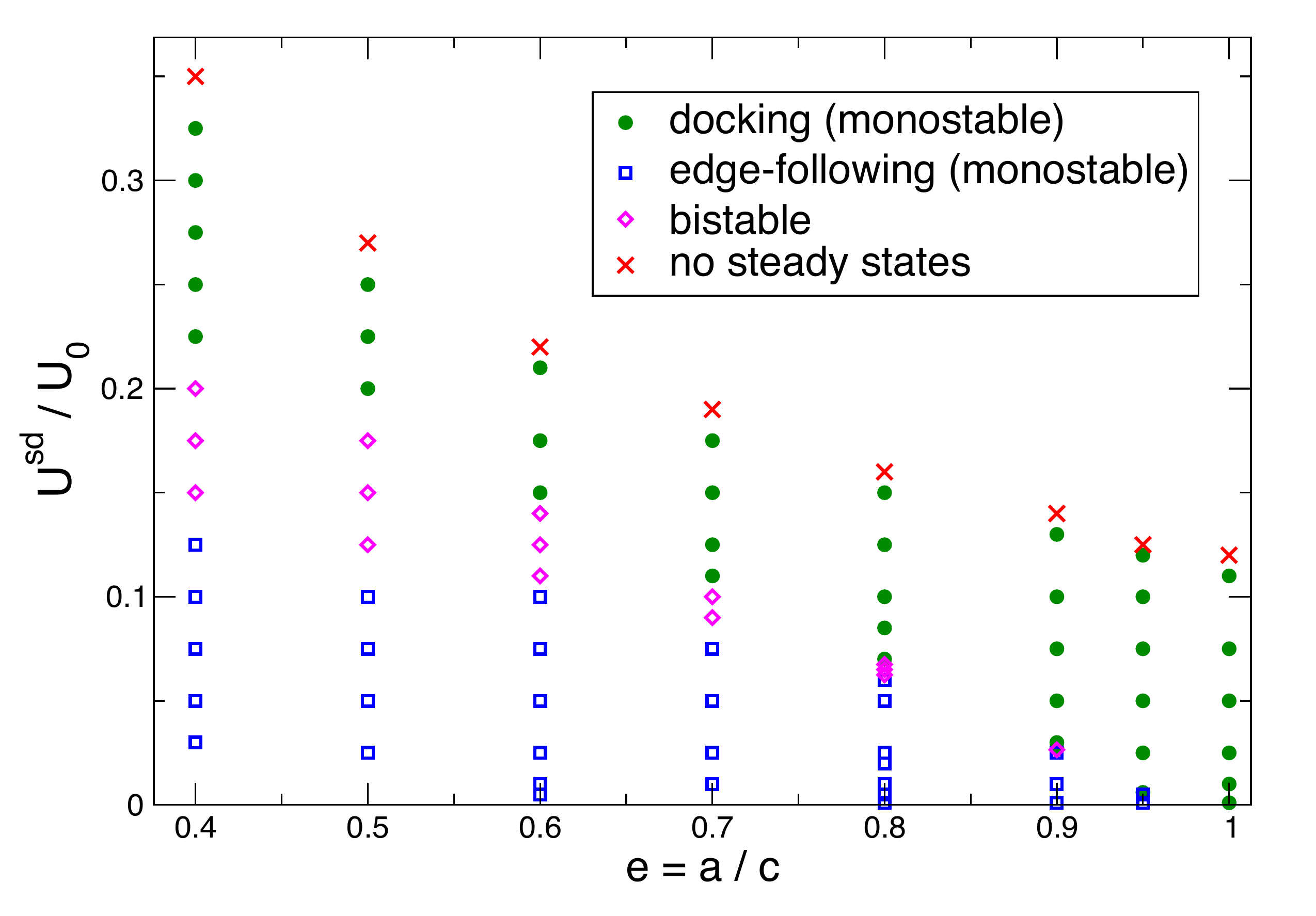}
\caption{\label{fig:phase_map} \WEU{Phase map showing the presence or absence of steady states for spherocylinders of various aspect ratios $e = a/c$ and self-propulsion speeds $U^{sd}/U_{0}$. The ratio of the particle height to the radius of the spherical end-caps is fixed to $h/a = 1.5$. Blue squares indicate that there is a monostable edge-following steady state at the corresponding combination of parameters $e$ and $U^{sd}$. Green circles indicate that there is a monostable docking state, and purple diamonds indicate that there are bistable edge-following and docking states.  Red crosses indicate that there are no steady states localized near the step.}}
\end{figure}

\WEU{We also briefly consider the effect of changing the particle aspect ratio  $e = a/c$. In order to address this issue, we consider spherocylinders with various aspect ratios located near the same chemical step as in Figs. \ref{fig:sphere_traj_portraits}-\ref{fig:phase_portrait_spherocyl}. We fix the ratio $h/a$, i.e., the ratio of the height of the particle center to the radius of the spherical end-caps of the particle, to $h/a = 1.5$. By decreasing $e = a/c$ from $e = 1$, the shape of the particle is smoothly deformed from spherical ($e = 1$) to quasi-spherical ($e \lesssim 1$) to rod-like ($e < 1$), while maintaining a constant thickness of the gap between the particle and the wall (as measured by the smallest dimension of the particle.) For each aspect ratio, we study particle behavior, in terms of the presence or absence of steady states, as we increase the self-propulsion velocity $U^{sd}$ from $U^{sd} = 0$. Our results are shown in Fig. \ref{fig:phase_map}. We find that even quasi-spherical particles ($e = 0.95$) have edge-following steady states, but these are restricted to a very small band near $U^{sd} = 0$. Decreasing the aspect ratio significantly increases the range of $U^{sd}$ over which edge-following states can be obtained. We also observe a region of bistability for $e \leq 0.9$, further extending the range of propulsion speeds for edge-following. For each aspect ratio considered, there are no steady states localized near the step for very large $U^{sd}$. However, decreasing the aspect ratio increases the threshold value of $U^{sd}$ for the disappearance of steady states.}

\MP{The results in Figs. \ref{fig:sphere_traj_portraits} - \ref{fig:phase_map} summarize the 
various phenomenology emerging from self-diffusiophoresis near chemically patterned walls. 
They clearly show that the shape of the particle is a relevant parameter for the dynamics.} Based on these observations, we seek to understand which new physical aspects are 
introduced by the non-spherical shape, as well as the mechanism by which they 
enrich the particle dynamics. We particularly focus on the rotation of the particle angle 
$\phi$. For a sphere, we note that a chemical step always drives rotation of the catalytic 
face towards the region of the substrate which is less repulsive to the solute. This fact 
can be clearly seen in the phase portraits of Fig. \ref{fig:sphere_traj_portraits}, 
because all trajectories in phase space always exhibit $\dot{\phi} \leq 0$, i.e., 
they always move from right to left in the diagrams. This observation suggests that 
there is a single dominant physical effect which drives the rotation of a 
sphere. For the spherocylinders, instead, the existence of edge-following attractors with 
$\phi$ between $\phi = 0$ and $\phi = \pi$ suggests that there might be multiple competing 
contributions to the angular velocity of the particle, and that these contributions balance 
for a particle in the edge-following state. 

\WEU{In order to understand} the interplay of the physical effects 
introduced by a non-spherical shape, in the following section we take an approximate analytical 
approach. We seek to identify and isolate the various competing contributions to translational 
and rotational velocities of a non-spherical particle. 

\section{Analytical results and discussion}

\subsection{Point-particle approach for the fluid velocity}

In order to obtain analytical expressions for $\mathbf{U}^{ws}$ and $\bm{\Omega}^{ws}$, 
we neglect the finite size of the particle under consideration, and treat it as a point-like object located at $\mathbf{x}_{p}$. Within this approximation, the fluid domain 
can be regarded as an infinite half-space 
$\lbrace \mathbf{x}_{f} \in \mathbb{R}^3 | z_f > 0 \rbrace$ bounded by a planar wall; 
the hydrodynamic boundary condition on the surface of the particle is neglected. There is a 
slip  velocity $\mathbf{v}_{s}(\mathbf{x}_{s})$ on the wall, where $\mathbf{x}_{s}$ 
denotes a position on the wall. The wall slip drives flow in the volume of the 
solution, and there are no external forces or torques applied to the system. We seek an 
expression for the fluid velocity $\mathbf{u}(\mathbf{x}_{f})$ at an \textit{arbitrary} point 
$\mathbf{x}_{f}$ in the fluid, i.e., $\mathbf{x}_{f}$  is not necessarily identical to 
$\mathbf{x}_{p}$.

We use the Lorentz reciprocal theorem, which relates the fluid stresses $(\bm{\sigma}, \bm{\sigma}')$ and velocity fields $(\mathbf{u}, \mathbf{u}')$ of two solutions of the 
Stokes equations which share the same geometry \cite{lorentz}:
\begin{equation}
\label{eq:reciprocal_thm}
\int \mathbf{u} \cdot \bm{\sigma}' \cdot \mathbf{\hat{n}} \, dS = \int \mathbf{u}' \cdot \bm{\sigma} \cdot \mathbf{\hat{n}} \, dS \,,
\end{equation}
where the integrals are performed over the wall with the surface normal 
$\mathbf{n} = \hat{\mathbf{z}}$ pointing into the fluid. We take the ``unprimed'' problem 
to be the one under consideration: the velocity on the wall is $\mathbf{u}(\mathbf{x}_{s}) = \mathbf{v}_{s}(\mathbf{x}_{s})$, and there are no external forces or torques. We consider three ``primed'' or auxiliary subproblems. For each of them, we take the wall to be a no-slip surface (i.e., $\mathbf{u}'(\mathbf{x}_{s}) = 0$). Additionally, there is an external force exerted on the fluid at point $\mathbf{x}_{f}$. The force has a magnitude $F_{0}$ and is oriented in $j$ direction, where $j \in \{x, y, z\}$ designates the auxiliary subproblem. The force creates a stress $F_0 \, \bm{\sigma}'^{(j)}(\mathbf{x}_{s}; \, \mathbf{x}_{f})$ 
in the fluid. Reformulating Eq. (\ref{eq:reciprocal_thm}) leads to \cite{uspal16}
\begin{equation}
\label{eq:velocity_in_fluid}
{u}_{j}(\mathbf{x}_{f}) = - \int \mathbf{v}_{s} (\mathbf{x}_{s}) \cdot \bm{\sigma}'^{(j)}(\mathbf{x}_{s}; \, \mathbf{x}_{f}) \cdot \mathbf{\hat{n}} \, dS \,.
\end{equation}
For each $j$ the integration kernel $\bm{\sigma}'^{(j)}(\mathbf{x}_{s}; \, \mathbf{x}_{f}) \cdot \mathbf{\hat{n}}$ is \cite{uspal16}:
\begin{equation}
\begin{split}
\label{eq:sigma_all_modes}
{\sigma}'^{(x)}_{iz}(\mathbf{x}_{s}; \, \mathbf{x}_{f}) &= - \frac{3 z_{f}}{2 \pi} \frac{r_i r_x}{r^{5}} \,,\\
{\sigma}'^{(y)}_{iz}(\mathbf{x}_{s}; \, \mathbf{x}_{f}) &= - \frac{3 z_{f}}{2 \pi} \frac{r_i r_y}{r^{5}} \,,~\mathrm{and}\\
{\sigma}'^{(z)}_{iz}(\mathbf{x}_{s}; \, \mathbf{x}_{f}) &=  \frac{3 z_{f}^{2}}{2 \pi} \frac{r_i}{r^{5}} \,,
\end{split}
\end{equation}
where $\mathbf{r} \equiv \mathbf{x}_{s} - \mathbf{x}_{f}$ and $r \equiv |\mathbf{r}|$. We 
note that although textbook statements of the reciprocal theorem are often restricted to 
Eq. (\ref{eq:reciprocal_thm}), an expression equivalent to Eq. (\ref{eq:velocity_in_fluid}) 
was already given in the original paper by Lorentz \cite{lorentz}.

Now we consider how the fluid velocity in Eq. (\ref{eq:velocity_in_fluid}) will affect the 
motion of the particle.  We apply Fax\'{e}n's Laws for the velocity of a point-like tracer 
particle in an ambient flow field \cite{spagnolie12}:
\begin{equation}
\label{eq:U_ws}
\mathbf{U}^{ws} = \mathbf{u}(\mathbf{x}_{p})
\end{equation}
and
\begin{equation}
\label{eq:Omega_ws}
\bm{\Omega}^{ws} = \frac{1}{2} \bm{\omega}(\mathbf{x}_{p}) + 
\Gamma \, \mathbf{\hat{d}} \times (\mathbf{E}(\mathbf{x}_{p}) \cdot \mathbf{\hat{d}}).
\end{equation}
The {vorticity} vector evaluated at the particle position is 
\begin{equation}
\bm{\omega}(\mathbf{x}_{p}) = \nabla \times \mathbf{u}(\mathbf{x}_{f}) \big|_{\mathbf{x}_{f} = \mathbf{x}_{p}},
\end{equation}
where the derivative is taken with respect to $\mathbf{x}_{f}$. The {strain rate tensor} 
evaluated at the particle position is 
\begin{equation}
\mathbf{E}(\mathbf{x}_{p}) = \frac{1}{2} \left( \nabla \mathbf{u}(\mathbf{x}_{f}) + \nabla^{T} \mathbf{u}(\mathbf{x}_{f}) \right) \big|_{\mathbf{x}_{f} = \mathbf{x}_{p}}.
\end{equation}
In Eq. (\ref{eq:Omega_ws}), the parameter $\Gamma$, with $0 \leq \Gamma \leq 1$, 
characterizes the elongation of the particle. It is zero for a sphere and close to unity 
for a long, slender object. For a prolate spheroid, there is an exact expression for 
$\Gamma$:
\begin{equation}
\label{eq:gamma_prolate}
\Gamma_{prolate} = \frac{1 - e^2}{1 + e^2}\,,
\end{equation}
where the spheroid aspect ratio $e = a/c$ is the ratio of the semi-minor axis length $a$ 
and the semi-major axis length $c$. For spherocylinders, there is no exact expression for 
$\Gamma$, but 
theoretical \cite{bretherton62} and experimental \cite{trevelyan51,snook14} studies have 
established that Eq. (\ref{eq:gamma_prolate}) holds as an excellent approximation if $e$ 
is replaced by an empirically determined effective aspect ratio $e_{eff} \approx 1.25 e$:
\begin{equation}
\label{eq:gamma_spherocyl}
\Gamma_{spherocyl} \approx \frac{1 - e_{eff}^2}{1 + e_{eff}^2}.
\end{equation}

In order to facilitate the application of Fax\'{e}n's Laws, by using Eq. (\ref{eq:velocity_in_fluid}) 
we have obtained
\begin{equation}
\label{eq:velocity_at_xp}
u_j(\mathbf{x}_{p}) =  - \int \mathbf{v}_{s} (\mathbf{x}_{s}) \cdot \bm{\sigma}'^{(j)}(\mathbf{x}_{s}; \, \mathbf{x}_{s}) \cdot \mathbf{\hat{n}} \, dS \,,
\end{equation}
\begin{equation}
\label{eq:vorticity_at_xp}
\omega_i(\mathbf{x}_{p}) =  - \int \left[ \epsilon_{ijk} \frac{\partial}{\partial x_{f,j}} (\mathbf{v}_{s}(\mathbf{x}_{s})  \cdot \bm{\sigma}'^{(k)}(\mathbf{x}_{s}; \mathbf{x}_{f}) \cdot \mathbf{\hat{n}} ) \right]_{\mathbf{x}_{f} = \mathbf{x}_{p}} \, dS \,,
\end{equation}
and
\begin{equation}
\label{eq:strain_rate_at_xp}
{E}_{ij}(\mathbf{x}_{p}) =  - \frac{1}{2} \int \left[ \frac{\partial}{\partial {x}_{f,i}} (\mathbf{v}_{s}(\mathbf{x}_{s})  \cdot \bm{\sigma}'^{(j)}(\mathbf{x}_{s}; \mathbf{x}_{f}) \cdot \mathbf{\hat{n}} ) +  \frac{\partial}{\partial {x}_{f,j}}(\mathbf{v}_{s}(\mathbf{x}_{s})  \cdot \bm{\sigma}'^{(i)} (\mathbf{x}_{s}; \mathbf{x}_{f}) \cdot \mathbf{\hat{n}} ) \right]_{\mathbf{x}_{f} = \mathbf{x}_{p}} \, dS,
\end{equation}
where $i$, $j$, and $k$ are elements of $\{x,y,z\}$, $x_{f,i}$ is the $i$ component of $\mathbf{x}_{f}$, 
and $\bm{\sigma}'^{(j)}(\mathbf{x}_{s}; \mathbf{x}_{f})$ is 
given by Eq. (\ref{eq:sigma_all_modes}).

We have now completely specified how to calculate the motion of a ``point-like'' particle 
advected by the flow in the volume of the solution. In turn, this flow is driven by 
the surface flow $\mathbf{v}_{s}(\mathbf{x}_{s})$ on a planar wall. At this point, for 
convenience, we introduce some new definitions and perform certain further reformulations. 
Concerning the rotational velocity, we define $\bm{\Omega}^{ws} \equiv 
\bm{\Omega}^{V} + \bm{\Omega}^{E}$, where we have distinguished the contributions from the 
vorticity (V) and the fluid strain rate (E), respectively. Given the assumed 
restriction of $\mathbf{\hat{d}}$ to the xy plane, we are interested only in the $\hat{z}$ 
component  $\Omega_{z}^{ws} = \Omega_{z}^{V} + \Omega^{E}_{z}$. 
\MP{By expanding the vector and scalar products in the last term of Eq. (\ref{eq:strain_rate_at_xp}), one 
arrives at}
\begin{equation}
\Omega^{E}_{z} =  \Gamma \,  \epsilon_{zjk} d_{j} E_{kl} d_{l},
\end{equation}
where
\begin{equation}
\begin{split}
 \epsilon_{zjk} d_{j} E_{kl} d_{l} &= \epsilon_{zxy} d_{x} E_{yl} d_{l} + \epsilon_{zyx} d_{y} E_{xl} d_{l} \\
 &=  d_{x} E_{yl} d_{l} - d_{y} E_{xl} d_{l} \\
 &= E_{yx} d_{x}^{2} + E_{yy} d_{x} d_{y} -  E_{xx} d_{x} d_{y} - E_{xy} d_{y}^{2}.
 \end{split}
\end{equation}
Since $E_{xy} = E_{yx}$, one has
\begin{equation}
\Omega^{E}_{z} = \Gamma E_{xy} (d_{x}^{2} - d_{y}^{2}) + \Gamma (E_{yy} -  E_{xx}) d_{x} d_{y}. 
\end{equation}
It is convenient to introduce
\begin{equation}
\label{eq:decomp_Ez}
\Omega^{E}_{z} \equiv \Omega^{E,{cross}}_{z}  + \Omega^{E,{diag}}_{z},
\end{equation}
where we have distinguished the contributions from the diagonal and the cross 
(i.e., off-diagonal) terms of the strain rate tensor.

\subsection{Point-particle approach for the solute field}

In order to be able to use Eqs. (\ref{eq:velocity_at_xp})-(\ref{eq:strain_rate_at_xp}), 
one needs analytical expressions for $\mathbf{v}_{s} = -b(\mathbf{x}_{s}) \nabla_{||} 
c(\mathbf{x})$. Here, we develop approximate, ``point-particle-like'' expressions for 
the solute number density field $c(\mathbf{x})$. 

The starting point is a multipole expansion for the solute number density field in 
\textit{unbounded solution}, i.e., with no confining boundaries. We consider the centroid 
of the particle (whether a sphere or a spherocylinder) to be located at the origin of a 
Cartesian coordinate system which is co-moving with the particle. The axis of symmetry 
of the particle is aligned with the $\hat{z}$ axis. The solute number density 
field is obtained, as defined in the beginning of Sect. II, by solving the Laplace 
equation subject to the boundary conditions noted there except for the one on the wall. 

The \textit{f}ree \textit{s}pace distribution $c^{fs}(\mathbf{x})$ of solute 
can be obtained numerically (e.g., by the BEM), or, for certain shapes of the 
particle, analytically. We seek to extract multipole expansion coefficients 
$\alpha_{i}$ from the result for $c^{fs}(\mathbf{x})$, where the coefficients 
$\alpha_{i}$ are determined by its expansion in terms of Legendre 
polynomials \cite{golestanian07}:
\begin{equation}
\label{eq:multipole_expansion}
c^{fs}(r, \theta) = c^{\infty} + \frac{R_{1}}{D} \sum_{l=0}^{\infty} \frac{\alpha_{l}}{l + 1} 
\left(\frac{R_{1}}{r}\right)^{l+1} P_{l}(\cos(\theta)), \; r \geq R_{1}\,.
\end{equation}
This expansion, expressed in terms of spherical coordinates, is taken to be valid in 
the fluid solution outside a sphere of radius $R_{1}$, where $2 R_{1}$ is the largest 
dimension of the particle. For a \textit{spherical} particle of radius $R$, this expansion 
is valid over the entire exterior region, i.e., $R_{1} = R$. 
Furthermore, if the solute number density field around the sphere obeys the boundary 
condition $[\mathbf{\hat{n}} \cdot \nabla c^{fs}(\mathbf{r})]|_{R,\theta} 
= -\kappa/D$ over the lower hemisphere (i.e., the catalyst covered area) of the 
sphere, and zero elsewhere on the surface of the sphere, the multipole coefficients 
$\alpha_{i}$ are straightforwardly obtained by using the orthogonality of the Legendre 
polynomials:
\begin{equation}
\label{eq:alphal_sphere}
\frac{\alpha_{l}}{\kappa} = \frac{2l + 1}{2} \int_{\pi/2}^{\pi} d\theta \sin(\theta) P_{l}
(\cos(\theta))\,.
\end{equation}
For instance, the monopole and dipole terms are $\alpha_{0} = \kappa/2$ and 
$\alpha_{1} = -3 \kappa/4$, respectively.

For other particle geometries, such as a spherocylinder, the expression for the 
coefficient $\alpha_{0}$ can be inferred from the following intuitive considerations. 
In the multipole expansion of Eq. \ref{eq:multipole_expansion}, the monopole term 
$\alpha_{0}$  encodes the net rate of solute production. For arbitrary geometries, we 
can write
\begin{equation}
\label{eq:alf_0}
\alpha_{0} = \frac{A_{catalyst}}{4 \pi R_{1}^{2}} \kappa,
\end{equation}
i.e., $\alpha_{0}$ is proportional to the ratio of catalyst area to the area of a sphere 
with diameter equal to $2 R_{1}$.  The higher order coefficients require detailed 
calculations.  In Appendix A, we discuss how these are obtained for an arbitrary shape of 
the particle from the numerically determined (e.g., via BEM) solute number density 
field $c^{fs}(\mathbf{x})$. Particularizing the general results to our specific 
spherocylinder geometry, we obtain (see Appendix A) $\alpha_0 = 0.2 \, \kappa$,  
$\alpha_1 = -\, 0.2262 \, \kappa$, etc. 

For a particle in the vicinity of the wall, the solute 
number density field can be approximated from the multipole expansion as follows. We 
restrict our consideration to the 
first two terms in the multipole expansion, and describe the particle as a point 
monopole plus a point dipole.\footnote{
\MP{We note that the strategy employed here is to keep the minimum number of terms in 
the multipole expansion based on which the approximate dynamics is in qualitative agreement 
with the numerical solution of the complete (exact) dynamics. In the sense of a perturbative 
approach, the approximation based on just the first two terms in the multipole expansion, combined with a 
point-particle solution of the hydrodynamics, is mathematically not justified when
$h/R$ is not much larger than 1, as studied here. Thus, the range of validity is determined 
``a posteriori'', similar to many other cases in which such approximations turn out to be accurate 
well beyond the mathematically rigorous limits of validity (see, e.g., Ref. \cite{spagnolie12}). For our system the approximate analysis seems to remain valid (at least 
quasi-quantitatively) down to very small values of $h/R-1$.}
} 
The direction of the dipole (defined to be from the 
point source to the point sink) is $\mathbf{\hat{d}}$. We place an image monopole and an 
image dipole below the wall at the position $(x_p, y_p, -h)$. The image monopole has the 
same strength $\alpha_0$ as the actual monopole. Likewise, the image dipole has the 
same strength $\alpha_1$ and orientation $\mathbf{\hat{d}}$ as the actual dipole 
(recalling that the dipole is parallel to the wall surface, since we assumed 
$\mathbf{\hat{d}} \cdot \hat{z} = 0$). Accordingly, one has
\begin{equation}
c(\mathbf{x}_{s}) \approx c^{mp}(\mathbf{x}_{s}) + c^{dp}(\mathbf{x}_{s}),
\end{equation}
where $c^{mp}(\mathbf{x}_{s})$ is the solute number density field due to the monopole 
at the position $\mathbf{x}_{s}$ on the wall, plus its image, and $c^{dp}(\mathbf{x}_{s})$ 
is the field due to the dipole at the position $\mathbf{x}_{s}$ on the wall, plus its 
image. For the monopole term, we obtain
\begin{equation}
\label{eq:c_mp}
c^{mp}(\mathbf{x}_{s}) = \frac{2 \alpha_{0} R_{1}^{2}}{D} 
\frac{1}{|\mathbf{x}_{s} - \mathbf{x}_{p}|}\,,
\end{equation}
and for the dipole term,
\begin{equation}
\label{eq:c_dp}
c^{dp}(\mathbf{x}_{s}) = - \frac{|\alpha_{1}| R_{1}^{3}}{D} 
\frac{\mathbf{\hat{d}} \cdot \mathbf{r}_{s}}{|\mathbf{x}_{s} - \mathbf{x}_{p}|^{3}},
\end{equation}
where $\mathbf{r}_{s} \equiv (x_s - x_p, y_s - y_p, 0)$. Consequently, the slip 
velocity $\mathbf{v}_{s}(\mathbf{x}_{s}) = - 
b(\mathbf{x}_{s}) \nabla_{||} c(\mathbf{x})|_{\mathbf{x}=\mathbf{x}_{s}}$ on the wall 
is decomposed as $\mathbf{v}_{s}(\mathbf{x}_{s}) \approx \mathbf{v}_{s}^{mp}(\mathbf{x}_{s}) + \mathbf{v}_{s}^{dp}(\mathbf{x}_{s})$.  The contribution of the monopole to the wall slip velocity is 
\begin{equation}
\label{eq:mp_vs}
\mathbf{v}_{s}^{mp} = \frac{2 \alpha_{0} R_{1}^{2}}{D} \frac{b_{w}(\mathbf{x}_{s}) \, \mathbf{r}_{s}}{|\mathbf{x}_{s} - \mathbf{x}_{p}|^{3}}\,,
\end{equation}
and the contribution of the dipole is
\begin{equation}
\label{eq:vs_dp}
\mathbf{v}_{s}^{dp} = \frac{b_w(\mathbf{x}_{s}) |\alpha_{1}| R_{1}^{3}}{D |\mathbf{x}_{s} - \mathbf{x}_{p}|^{3}} \,\; \mathbf{\hat{d}} \cdot  \left( \mathcal{I} - \frac{3 \mathbf{r}_{s} \mathbf{r}_{s}}{|\mathbf{x}_{s} - \mathbf{x}_{p}|^{2}} \right) .
\end{equation}

\subsection{Monopole contribution to the particle velocity}

The combination of Eqs. (\ref{eq:velocity_in_fluid}), (\ref{eq:sigma_all_modes}), and 
(\ref{eq:mp_vs}) yields the monopole contributions to the fluid velocity at an arbitrary 
point $\mathbf{x}_{f}$ above a chemically patterned wall:

\begin{equation}
{u}_{x}^{mp}(\mathbf{x}_{f}) =  \frac{3 h}{2 \pi} \frac{2 \alpha_{0} R^{2}}{D} \int \frac{b_{w}(\mathbf{x}_{s})(x_{s}-x_{f})[(x_{s}-x_{f})(x_{s}-x_{p}) + (y_{s} - y_{f})(y_{s}-y_{p})]}{((x_{f}-x_{s})^2 + (y_{f} - y_{s})^{2} + z_{f}^{2})^{5/2} ((x_{p} - x_{s})^{2} + (y_{p} - y_{s})^{2} + h^{2}))^{3/2}} \, dS
\end{equation}
and
\begin{equation}
{u}_{y}^{mp}(\mathbf{x}_{f}) =  \frac{3 h}{2 \pi} \frac{2 \alpha_{0} R^{2}}{D} \int \frac{b_{w}(\mathbf{x}_{s})(y_{s}-y_{f})[(x_{s}-x_{f})(x_{s}-x_{p}) + (y_{s} - y_{f})(y_{s}-y_{p})]}{((x_{f}-x_{s})^2 + (y_{f} - y_{s})^{2} + z_{f}^{2})^{5/2} ((x_{p} - x_{s})^{2} + (y_{p} - y_{s})^{2} + h^{2}))^{3/2}} \, dS.
\end{equation}

Now we consider the derivatives of the fluid velocity. In order to obtain $E_{xy}^{mp}(\mathbf{x}_{p})$, we calculate $\frac{1}{2} \left( {\partial u_{x}^{mp}}/{\partial y_{f}} + {\partial u_{y}^{mp}}/{\partial x_{f}} \right)\big|_{\mathbf{x}_{f} = \mathbf{x}_{p}}$. The two partial derivatives are identical, yielding
\begin{equation}
E_{xy}^{mp}(\mathbf{x}_{p}) =  \frac{3 h}{2 \pi} \frac{2 \alpha_{0} R^{2}}{D} \int \frac{b_{w}(\mathbf{x}_{s}) (x_{p} - x_{s})(y_{p} - y_{s}) [4 \, ((x_{p} - x_{s})^2 + (y_{p} - y_{s})^2) - h^{2}]}{((x_{p} - x_{s})^{2} + (y_{p} - y_{s})^2 + h^{2})^{5}} \, dS.
\end{equation}

Analogously one obtains $E_{xx}^{mp}(\mathbf{x}_{p})$ and $E_{yy}^{mp}(\mathbf{x}_{p})$:
\begin{equation}
\begin{split}
& E_{xx}^{mp}(\mathbf{x}_{p}) = \frac{3 h}{2 \pi} \frac{2 \alpha_{0} R^{2}}{D} \\ & \, \, 
\times \int \frac{b_{w}(\mathbf{x}_{s}) [3 (x_p-x_s)^4 + 2(x_p-x_s)^2(y_p-y_s)^2 - (y_p-y_s)^4-(2(x_p-x_s)^2+(y_p-y_s)^2)h^2]}{((x_p - x_s)^2 + (y_p - y_s)^2 + h^2)^5} \, dS
\end{split} 
\end{equation}
and
\begin{equation}
\begin{split}
& E_{yy}^{mp}(\mathbf{x}_{p}) = -\frac{3 h}{2 \pi} \frac{2 \alpha_{0} R^{2}}{D} \\ & \;\; 
\times  \int \frac{b_{w}(\mathbf{x}_{s}) [((x_p-x_s)^2 - 3(y_p-y_s)^2)((x_p-x_s)^2+(y_p-y_s)^2)+((x_p-x_s)^2+2(y_p-y_s)^2)h^2]}{((x_{p} - x_{s})^{2} + (y_{p} - y_{s})^2 + h^{2})^{5}} \, dS\,.
\end{split} 
\end{equation}

The monopolar contribution to the vorticity component $\omega_{z}$ vanishes, because the two partial derivatives in $\omega_z = \left( {\partial u_{x}^{mp}}/{\partial y_{f}} - {\partial u_{y}^{mp}}/{\partial x_{f}} \right)\big|_{\mathbf{x}_{f} = \mathbf{x}_{p}}$ are identical so that
\begin{equation}
\omega^{mp}_{z} = 0.
\end{equation}

If the pattern exhibits translational symmetry in the $\hat{\mathbf{y}}$ direction, 
such that $b(\mathbf{x}_{s}) = b(x_{s})$, the particle velocity does not depend on $y_p$.  
Accordingly, in calculating the velocity, one can set $y_p = 0$ without loss of generality. 
Concerning $E_{xy}^{mp}({x}_{p})$, the resultant expression involves an integral over an 
odd function of $y_s$, and hence it is zero:
\begin{equation}
\label{eq:Exx_y_sym}
E_{xy}^{mp}({x}_{p}) = - \frac{3 h}{2 \pi} \frac{2 \alpha_{0} R^{2}}{D} \int dx_{s} \int dy_{s} \,  \frac{b_{w}({x}_{s}) (x_{p} - x_{s}) y_{s} [4 \, ((x_{p} - x_{s})^2 +  y_{s}^2) - h^{2}]}{((x_{p} - x_{s})^{2} + y_{s}^2 + h^{2})^{5}}  = 0.
\end{equation}

Now we consider the specific case of a chemical step by substituting 
Eq. (\ref{eq:step_surface_mob}) for $b_{w}(\mathbf{x}_{s})$.  According to 
Eq. (\ref{eq:Exx_y_sym}), $E_{xy}^{mp}$ vanishes. This leaves four piecewise integrals to 
be performed: the contributions of the left ($x_s < 0$) and of the right ($x_s \,>\, 0$) 
side to $E_{xx}^{mp}$ and $E_{yy}^{mp}$. After straightforward but cumbersome analysis, we 
obtain the contribution of the monopole to particle rotation:
\begin{equation}
\label{eq:omega_z_mp}
\Omega_{z}^{mp} = -\frac{3 \alpha_{0} \Gamma R_{1}^{2}}{128 D} \, (b_{w}^{l} - b_{w}^{r}) \, \frac{h (32 x_{p}^{3} + 7 x_{p} h^{2})}{(x_{p}^{2} + h^{2})^{7/2}} \sin(\phi) \cos(\phi)\,.
\end{equation}

\subsection{Dipole contribution to the particle velocity}

By combining Eqs. (\ref{eq:velocity_in_fluid}), (\ref{eq:sigma_all_modes}), and 
(\ref{eq:vs_dp}) one finds the dipolar contributions to the fluid velocity at 
$\mathbf{x}_{f}$:
\begin{equation}
\begin{split}
\label{eq:ux_dp}
u_{x}^{dp}(\mathbf{x}_{f}) = \frac{3 h}{2 \pi} \frac{|\alpha_{1}| R_{1}^{3}}{D} \left[\int   \frac{  b_{w}(\mathbf{x}_{s})(x_s - x_f) ((x_s-x_f) d_x + (y_s - y_f) d_y)}{((x_f - x_s)^2 + (y_f - y_s)^2 + h^2)^{5/2} ((x_p - x_s)^2 + (y_p - y_s)^2  h^2)^{3/2}} \, dS \, + \, \right. \\ \left. 
- \int \frac{3 b_{w}(\mathbf{x}_{s}) (x_s-x_f)[(x_s-x_p) d_x + (y_s - y_p) d_y] [(x_s - x_f)(x_s - x_p) + (y_s - y_f)(y_s - y_p)]}{((x_f - x_s)^2 + (y_f - y_s)^2 + h^2)^{5/2} ((x_p - x_s)^2 + (y_p - y_s)^2 + h^2)^{5/2}} \, dS \right]
\end{split}
\end{equation}
and
\begin{equation}
\begin{split}
\label{eq:uy_dp}
u_{y}^{dp}(\mathbf{x}_{f}) = \frac{3 h}{2 \pi} \frac{|\alpha_{1}| R_{1}^{3}}{D} \left[\int   \frac{  b_{w}(\mathbf{x}_{s})(y_s - y_f) ((x_s-x_f) d_x + (y_s - y_f) d_y)}{((x_f - x_s)^2 + (y_f - y_s)^2 + h^2)^{5/2} ((x_p - x_s)^2 + (y_p - y_s)^2  h^2)^{3/2}} \, dS \, + \, \right. \\ \left. 
- \int \frac{3 b_{w}(\mathbf{x}_{s}) (y_s-y_f)[(x_s-x_p) d_x + (y_s - y_p) d_y] [(x_s - x_f)(x_s - x_p) + (y_s - y_f)(y_s - y_p)]}{((x_f - x_s)^2 + (y_f - y_s)^2 + h^2)^{5/2} ((x_p - x_s)^2 + (y_p - y_s)^2 + h^2)^{5/2}} \, dS \right],
\end{split}
\end{equation}
where, in each equation, the two integrals correspond to the two terms in the parentheses in 
Eq. (\ref{eq:vs_dp}). As with the monopole contributions,  one can construct surface 
integrals for the vorticity and the strain rate tensor associated with the dipole by 
taking partial derivatives of Eqs. (\ref{eq:ux_dp}) and (\ref{eq:uy_dp}). We omit these 
lengthy intermediate expressions, and provide only the final results, which are, somewhat 
surprisingly, rather compact expressions. The contribution of the dipole to rotation 
via vorticity (see the text below Eq. (\ref{eq:strain_rate_at_xp})) is 
\begin{equation}
\label{eq:omega_z_v_dp}
\Omega_{z}^{V,dp} = - \frac{3 |\alpha_{1}| R_{1}^{3} }{64 D} \, (b_{w}^{l} - b_{w}^{r}) \, 
\frac{h}{(h^2 + x_p^{2})^{5/2}} \sin(\phi)\,.
\end{equation}
The dipolar contributions to rotation via the rate of strain (see Eq. 
(\ref{eq:decomp_Ez}) and the text below Eq. (\ref{eq:strain_rate_at_xp})) are
\begin{equation}
\label{eq:omega_z_E_cross_dp}
\Omega_{z}^{E,{cross},dp}(\mathbf{x}_{p}) = \frac{3 |\alpha_{1}| R_{1}^{3} \Gamma }{512 D} (b_{w}^{l} - b_{w}^{r}) \,  \frac{h (22 x_p^2 + 7 h^{2})}{(x_p^2 + h^2)^{7/2}}  \sin(\phi) \, [\cos^{2}(\phi) - \sin^{2}(\phi)]
\end{equation}
and
\begin{equation}
\label{eq:omega_z_E_diag_dp}
\Omega_{z}^{E,{diag},dp}(\mathbf{x}_{p}) = - \frac{15 |\alpha_{1}|R_{1}^{3}  \Gamma }{512 D}  (b_{w}^{l} - b_{w}^{r}) \, \frac{h (20 x_p^4 + 2 x_{p}^{2} h^{2} + 3 h^{4})}{(x_p^2 + h^2)^{9/2}} \, \cos^{2}(\phi) \sin(\phi)\,.
\end{equation}
\begin{figure}[!htb]
\includegraphics[scale=0.9]{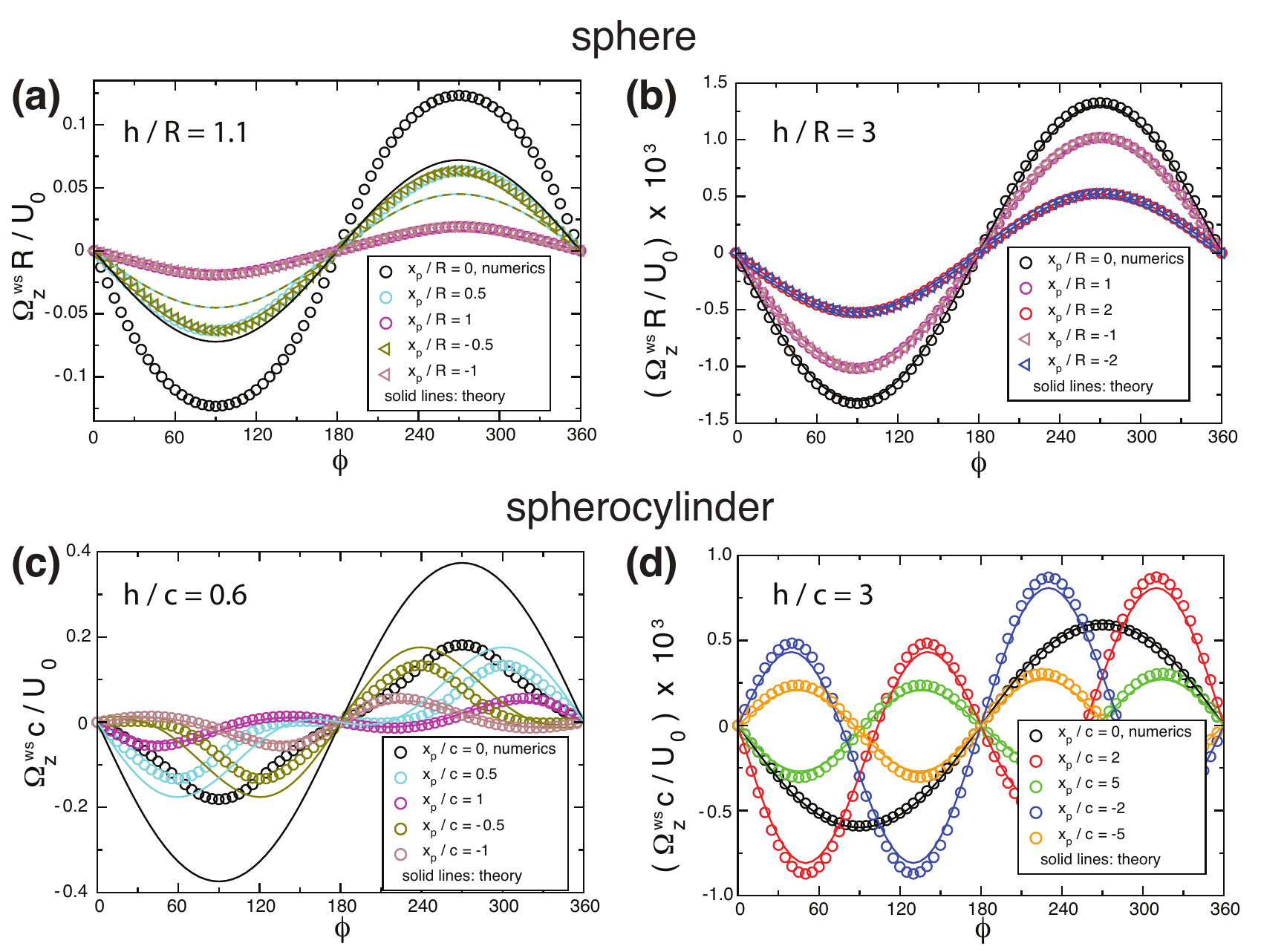}
\caption{\label{fig:sphere_spherocyl_omegaZ} (a) and (b) Angular velocity $\Omega_{z}^{ws}$ of 
a half-covered sphere due to chemi-osmotic slip at the wall as a function of orientation 
$\phi$ for several lateral distances $x_{p}$ from a chemical step. The step is characterized 
by $b_{w}^{r}/b_{w}^{l} = 4$ and $b_{w}^{l} < 0$ and is located at $x_{p} = 0$. In (a), 
the center of the sphere is at a height $h/R = 1.1$ above the surface. In (b), the height is $h/R = 3$.  (c) and (d) Angular velocity of a half-covered spherocylinder as a function of orientation $\phi$ for several lateral distances from the same chemical step. In (c), the centroid of the spherocylinder is at height $h/c = 0.6$ above the surface, while 
in (d) it is at height $h/c = 3$, where $2 c$ is the tip-to-tip length of the 
spherocylinder. In (a) the theoretical and numerical data for $x_p/R = \pm 1$ \textit{de facto} coincide. In (b) the theoretical and numerical data for $x_p/R = \pm 1$ \textit{de facto} coincide, as do the theoretical and numerical data for $\pm 2$. }
\end{figure}

\subsection{Comparison of analytical and numerical results}

Finally, by combining Eqs. (\ref{eq:omega_z_mp}), (\ref{eq:omega_z_v_dp}), 
(\ref{eq:omega_z_E_cross_dp}), and (\ref{eq:omega_z_E_diag_dp}) one obtains an approximation 
expression for the rotation of a particle in the vicinity of a patterned wall:
\begin{equation}
\label{eq:omega_Z_ws_all_terms}
\Omega_{z}^{ws}(\mathbf{x}_{p}) \approx \Omega_{z}^{mp}(\mathbf{x}_{p}) + \Omega_{z}^{V,dp}(\mathbf{x}_{p}) + \Omega_{z}^{E,{cross},dp}(\mathbf{x}_{p}) + \Omega_{z}^{E,{diag},dp}(\mathbf{x}_{p}).
\end{equation}
We first consider the rotation of a sphere near a chemical step. For this shape of the 
particle, $\Omega_{z}^{mp}(\mathbf{x}_{p})$, $\Omega_{z}^{E,{cross},dp}(\mathbf{x}_{p})$, 
and $\Omega_{z}^{E,{diag},dp}(\mathbf{x}_{p})$ all vanish, because they originate from the 
strain rate tensor. The remaining term, $\Omega_{z}^{V,dp}(\mathbf{x}_{p})$, is a simple 
sinusoid (Eq. \ref{eq:omega_z_v_dp}), and hence crosses zero only at $\phi = 0$, 
$\phi = \pi$, and $\phi = 2 \pi$. Therefore, the only possible stable orientations of the 
sphere are docking states (i.e., pointing towards or away from the step). In 
Figs. \ref{fig:sphere_spherocyl_omegaZ}(a) and (b) we compare this expression with 
\MP{the numerical solution of the exact dynamics, which is} obtained by BEM. In this figure, (a) 
corresponds to $h/R = 1.1$ and (b) to $h/R = 3$, and $\Omega_{z}^{ws}$ is plotted as a 
function of $\phi$ for various lateral distances $x_{p}/R$ from the step. We obtain excellent 
agreement for $h/R = 3$, for which we expect the truncation of the multipole expansion to the 
monopole and dipole terms to be a good approximation. Closer to the surface ($h/R = 1.1$), we 
obtain qualitative, but only semi-quantitative agreement. In particular, we note that while 
the numerical results show a stronger rotation by the surface than predicted by the 
approximate expressions, they retain the theoretically predicted (see above) sinusoidal 
dependence on $\phi$.

For the spherocylinder, all terms in Eq. (\ref{eq:omega_Z_ws_all_terms}) contribute. 
The resultant expression for $\Omega_{z}^{ws}$ is complicated, and can cross $\Omega_{z}^{ws} 
= 0$ between $\phi = 0$ and $\phi = \pi$. This more complicated dependence enables the 
occurrence of the edge-following state. In Figs. \ref{fig:sphere_spherocyl_omegaZ}(c) and 
(d), we show the theoretical and numerical results for a spherocylinder at $h/c = 0.6$ and 
$h/c = 3$, respectively. For $h/c = 3$, we again obtain excellent agreement. 
For $h/c = 0.6$, there is no \textit{a priori} reason to expect agreement, because our 
mapping of the particle solute field to the multipole expansion was performed for distances 
from the particle centroid larger than $R_{1} = c$. Nevertheless, the agreement remains 
surprisingly good. The strongest deviation occurs for $x_p = 0$, for which the numerical 
results exhibit a more complicated functional dependence on $\phi$ than the sinusoidal one 
predicted theoretically. Note that for $x_p/c = 1$ (brown circles and brown curve) and 
$x_p/c = -1$ (purple circles and purple curve), the disagreement between theory and 
numerics can hardly be distinguished by eye.

\section{Conclusions}

We have studied the effect of particle shape on the motion of a catalytically active particle 
near a chemically patterned surface. The active particle locally induces chemi-osmotic flows 
on the surface, which, in turn, drive flows in the volume of the solution 
that couple back to the particle. This chemi-osmotic contribution to particle motion is sensitive 
to the material identity of the surface, and it competes with the inherent 
self-diffusiophoretic motion of the particle. We have restricted our focus to 
``inert-forward'' particles which tend to move away from their catalytic caps via 
self-diffusiophoresis.  As first shown in previous work, near a chemical step 
``inert-forward'' spherical particles exhibit a rich phenomenology \cite{uspal16}. A sphere 
can pass over the step, or stably ``dock'' at the edge between the two materials, depending 
on the system parameters and the initial position and orientation of the particle. 
A docked particle remains motionless with its axis of symmetry aligned perpendicular to 
the edge of the step.  A particle which passes over the step can exhibit 
significant deviations of its orientation from the angle of approach.  If the particle is 
non-spherical, such as a spherocylinder, this phenomenology is significantly enriched. In addition 
to the docking behavior, a rod-like particle can stably follow the edge of the step, with its axis 
of symmetry tilted slightly away from alignment with the edge.  This ``edge-following'' state can 
coexist with the docking state for a broad range of parameters. Additionally, if the surface has a 
narrow chemical stripe (i.e., two neighboring step-like interfaces), the edge-following state is 
transformed into a stable ``stripe-following'' 
state in which the particle is located at the stripe center and is perfectly aligned 
with the stripe. 

These observations motivated the development of an approximate analytical approach in order to 
understand the underlying physical mechanisms. A central insight of our approach is a distinction 
between the orientational response of a particle to vorticity and to the fluid strain rate. All 
particles are driven to rotate by fluid vorticity, but only elongated (i.e., non-spherical) 
particles respond to the hydrodynamic strain rate. Our analytical expressions demonstrate that 
this significantly more complicated orientational behavior is the basis of the enriched 
phenomenology observed for spherocylinders. Therefore, our findings are not specific to 
spherocylinders, but generalize to other non-spherical shapes (e.g., prolate 
spheroids \cite{popescu10}), as detailed numerical calculations (omitted here) confirm. Our 
numerical and analytical calculations have considered a self-diffusiophoretic particle with a 
monopole term, i.e., a net producer of product molecules. However, our theoretical framework can 
also be applied to confined self-electrophoretic particles\MP{, under the additional assumption of the presence of
a thin Debye layer required by a mapping to an osmotic slip velocity (which has been used in, e.g., 
Ref. \cite{liu16}),} with appropriate renaming of the variables: $c(\mathbf{x})$ as an electrical 
potential $\psi(\mathbf{x})$, the surface mobility $b(\mathbf{x}_{s})$ as a quantity proportional 
to the zeta potential $\zeta(\mathbf{x}_{s})$, etc. \cite{anderson89,golestanian07,chiang14,liu16}.
A self-electrophoretic particle would have no monopole term ($\alpha_{0} = 0$), but 
otherwise our findings would carry over to this new setting.

\MP{While the focus of the current work is on the deterministic dynamics, the effects of 
moderate thermal fluctuations can be assessed, to a certain extent, from the dynamical stability of 
the corresponding fixed points. The linear stability analysis of the deterministic equations is 
straightforward, requiring only evaluations of the Jacobians at the fixed points. Similarly with 
the observations in our previous work \cite{uspal15}, the attractors of the dynamics are 
``stiff'', which prevents large excursions from the vicinity of the attractor in the phase plane. For instance, following the procedure of Ref. \cite{uspal15}, we obtain a longest relaxation time of $3.7\;\textrm{s}$ for the docking state in Fig. \ref{fig:sphere_traj_portraits}(c), which compares favorably with the characteristic rotational diffusion time of $D_{r}^{-1} = 95\;\textrm{s}$. However, one often encounters, as in Fig. \ref{fig:stripe_figure}(a), a combination of a fast dynamics along one coordinate (the translation) with a (very) slow dynamics along the other coordinate (i..e, the rotation of the particle axis). This  implies that the small excursions in the phase plane may actually translate into extended periods 
of time that a particle, subject to stochastic dynamics, spends outside the fixed point (see also, 
e.g., Ref. \cite{mozaffari17}).
}

Our findings open new possibilities for programming the motion of catalytically active particles 
in lab-on-a-chip devices. The edge-following and stripe-following states can be used to direct 
rod-like particles from one location to another. If the ratio $U^{sd}/U_{0}$ (characterizing the 
relative contributions of self-diffusiophoresis and chemi-osmosis to motility) is sensitive to 
light or some other stimulus, a rod-like particle could be switched between edge-following and 
docking states on demand. This would be the case, for instance, if light could be used to evoke 
changes in the surface properties of either the particle or the wall. We note that we had 
previously predicted stripe-following behavior for spheres, but only for the ``catalyst-forward'' 
direction of motion, which is much less commonly seen in experiments. Additionally, rod-like 
particles quite naturally realize our assumption of quasi-2D motion, in which the particle is 
restricted to have a constant height above the wall and an orientation vector that always remains 
within a plane parallel to the wall. \WEU{Our findings concerning the role of the particle aspect ratio show that for $e < 0.85$ the edge-following states exist over a broad range of self-propulsion speeds.}

Further work could also investigate more complicated shapes. L-shaped particles which 
are restricted to quasi-2D motion have a well-defined chirality \cite{kummel13}. The effect 
of fore-aft asymmetry (e.g., for an asymmetric dumbbell with a large ``head'') is an open 
question \cite{valadares10,uspal13}. Additionally, one could consider shapes with a more 
complicated topology, such as ring or toroidal shapes \cite{schmieding17,singh13}.

\acknowledgments We thank C. Pozrikidis for making the \texttt{BEMLIB} library freely available 
\cite{pozrikidis02} \WEU{and an anonymous referee for suggesting Fig. \ref{fig:phase_map}}. W.E.U., M.T., and M.N.P. acknowledge financial support from the German Science Foundation (DFG), grant no. TA 959/1-1. \WEU{M.T. acknowledges financial support from the Portuguese Foundation for Science and Technology (FCT) under Contracts no. IF/00322/2015.}

\section*{Appendix A: numerical calculation of multipole coefficients}

For many geometries of interest, such as the spherocylinder, there is no suitable system of 
coordinates in terms of which the Laplace equation is separable. For these geometries, 
instead a numerical method, such as BEM, can be used to obtain $c(\mathbf{r})$. We 
briefly outline how to obtain the multipole coefficients from a known solute density 
$c(\mathbf{r})$ surrounding an axisymmetric particle in unbounded solution.

Any solution $c(\mathbf{r})$ of the Laplace equation can be expressed as an integral 
over the domain boundaries \cite{pozrikidis02}:
\begin{equation}
\label{eq:boundary_integral_c}
c(\mathbf{r}') = c_{\infty} - 
\int_{\cal D} G(\mathbf{r}', \mathbf{r}) [\mathbf{\hat{n}} \cdot \nabla c(\mathbf{r})] dS + \int_{\cal D} c(\mathbf{r}) [ \mathbf{\hat{n}} \cdot \nabla_{\mathbf{r}} G( \mathbf{r}', \mathbf{r}) ] dS.
\end{equation}
Here, ${\cal D}$ is the particle surface, $\mathbf{r}'$ is a point of interest in the 
fluid domain, and the local surface normal $\mathbf{\hat{n}}$ is defined to point from the particle surface into the fluid solution. The Green's function $G(\mathbf{r}, \mathbf{r}')$ is
\begin{equation}
\label{eq:fs_gf}
G(\mathbf{r}, \mathbf{r}') \equiv \frac{1}{4 \pi |\mathbf{r} - \mathbf{r}'|}.
\end{equation}
Equations (\ref{eq:boundary_integral_c}) and (\ref{eq:fs_gf}) indicate that one can 
interpret the solution ${c}(\mathbf{r})$ as arising from a distribution of point 
sources (monopoles) and point source - point sink pairs (dipoles) over the particle 
surface which, respectively, correspond to the first and the second integrals 
in Eq. (\ref{eq:boundary_integral_c}).

Now we replace the left hand side of Eq. (\ref{eq:boundary_integral_c}) by 
Eq. (\ref{eq:multipole_expansion}):
\begin{equation}
\frac{R_{1}}{D} \sum_{l=0}^{\infty} \frac{\alpha_{l}}{l + 1} 
\left(\frac{R_{1}}{r'}\right)^{l+1} P_{l}(\cos(\theta')) = - \int_{\cal D} G(\mathbf{r}', 
\mathbf{r}) [\mathbf{\hat{n}} \cdot \nabla c(\mathbf{r})] dS + \int_{\cal D} c(\mathbf{r}) 
[ \mathbf{\hat{n}} \cdot \nabla_{\mathbf{r}} G( \mathbf{r}', \mathbf{r}) ] dS.
\end{equation}
We denote the spherical surface defined by $r' = R_{1}$ as ${{\cal D}_1}$. 
We multiply both sides by $P_{n}(\cos(\theta'))$ and integrate them as 
$\int_{{\cal D}_1} dS' $. Due to the orthogonality of the Legendre polynomials, one has
\begin{multline}
\label{eq:alpha_n_bie}
\frac{2 \pi R_{1}}{D}  \frac{\alpha_{n}}{n + 1} \frac{2}{2n + 1} = 
- \int_{{\cal D}} dS \, [\mathbf{\hat{n}} \cdot \nabla c(\mathbf{r})] \int_{{\cal D}_1} dS' \, G(\mathbf{r}', \mathbf{r}) P_{n}(\cos(\theta')) \\ 
+ \int_{{\cal D}} dS \, c(\mathbf{r}) \, \mathbf{\hat{n}} \cdot \nabla_{\mathbf{r}} \int_{{\cal D}_1} dS' \, G( \mathbf{r}', \mathbf{r}) P_{n}(\cos(\theta'))\,. 
\end{multline}
In the next step we expand the Green's function in terms of spherical harmonics:
\begin{equation}
G(\mathbf{r}', \mathbf{r}) = \sum_{l=0}^{\infty} \frac{1}{2l + 1} \sum_{m=-l}^{l} (-1)^{m} 
\frac{r^{l}}{r'^{l+1}} Y_{l,-m}(\theta, \phi) Y_{l,m}(\theta', \phi')\,.
\end{equation}
This leads to
\begin{equation}
\label{eq:integr_G}
\begin{split}
\int_{{\cal D}_1} dS' \, G(\mathbf{r}',\mathbf{r}) P_{n}(\cos(\theta')) 
& = 
\sum_{l=0}^{\infty} \frac{1}{2l + 1} \sum_{m=-l}^{l} \frac{r^{l}}{R_{1}^{l+1}}  
Y_{l,-m}(\theta, \phi) \int_{{\cal D}_1} dS'\, Y_{l,m}(\theta', \phi') 
\, P_{n}(\cos(\theta'))  \\
& = \frac{1}{2n + 1} \frac{r^{n}}{R_{1}^{n+1}} Y_{n,0}(\theta, \phi) 
\sqrt{\frac{4 \pi}{2n + 1}} \\
&= \frac{1}{2n + 1} \frac{r^{n}}{R_{1}^{n+1}} P_{n}(\cos(\theta))\,,
\end{split}
\end{equation}
where upon computing the integral on the right hand side above we have used 
$P_{n}(\cos(\theta')) = \sqrt{\frac{2n+1}{4 \pi}} ~ Y_{n,0}(\theta',\phi')$ and the 
orthogonality of the spherical harmonics.

Substituting Eq. (\ref{eq:integr_G}) into Eq. (\ref{eq:alpha_n_bie}) renders
\begin{equation}
\label{eq:alpha_n_bie_2}
\frac{4 \pi R_{1}}{D}  \frac{\alpha_{n}}{n + 1}  = - \int_{{\cal D}} dS 
\, [\mathbf{\hat{n}} \cdot \nabla c(\mathbf{r})] \frac{r^{n}}{R_{1}^{n+1}} P_{n}(\cos(\theta))  
+ 
\int_{{\cal D}} dS \, c(\mathbf{r}) \, \mathbf{\hat{n}} \cdot \nabla_{\mathbf{r}} 
\left[ \frac{r^{n}}{R_{1}^{n+1}} P_{n}(\cos(\theta)) \right].
\end{equation}

For $\alpha_{0}$, only the first integral contributes:
\begin{equation}
\alpha_{0} = - \frac{D}{4 \pi R_{1}^{2}} \int_{D} dS \, [\mathbf{\hat{n}} 
\cdot \nabla c(\mathbf{r})].
\end{equation}
Recalling that $[\mathbf{\hat{n}} \cdot \nabla c(\mathbf{r})] = -\kappa/D$ over the 
catalyst covered area $A_{catalyst}$ and zero elsewhere (see the first paragraph in 
Sect. II), we have
\begin{equation}
\alpha_{0} = \frac{A_{catalyst}}{4 \pi R_{1}^{2}} \kappa,
\end{equation}
which we obtained earlier by simple arguments (see the text preceding 
Eq.(\ref{eq:alf_0})). 

The dipole term $\alpha_{1}$ has contributions from both integrals in 
Eq. (\ref{eq:alpha_n_bie_2}):
\begin{equation}
\frac{2 \pi R_{1}}{D} \alpha_{1}  = - \int_{D} dS \, [\mathbf{\hat{n}} \cdot \nabla c(\mathbf{r})] \frac{r}{R_{1}^{2}} P_{1}(\cos(\theta))  + \int_{D} dS \, c(\mathbf{r}) \, \mathbf{\hat{n}} \cdot \nabla_{\mathbf{r}} \left[ \frac{r}{R_{1}^{2}} P_{1}(\cos(\theta)) \right].
\end{equation}
Using $r \, P_{1}(\cos(\theta)) = z$, one finds 
\begin{equation}
\label{eq:alpha1_BIE}
\frac{2 \pi R_{1}^{3}}{D} \alpha_{1}  = - \int_{D} dS \, [\mathbf{\hat{n}} \cdot \nabla c(\mathbf{r})] z  + \int_{D} dS \, c(\mathbf{r}) \, \mathbf{\hat{n}} \cdot \hat{z}\,.
\end{equation}
One can readily verify that the solution for a sphere 
(Eq. (\ref{eq:multipole_expansion})) satisfies Eq. (\ref{eq:alpha1_BIE}).

\bibliography{rods_citations}

\end{document}